\newcommand{\beq}{\begin{equation}}
\newcommand{\eeq}{\end{equation}}

\newcommand{\f}{\frac}

\newcommand{\po}{\left(}
\newcommand{\pc}{\right)}
\newcommand{\ls}{\thinspace}
\newcommand{\lls}{\ls\ls}
\newcommand{\bs}{\thickspace}
\newcommand{\bbs}{\bs\bs}

\newcommand{\ep}{\epsilon}

\newcommand{\itR}{\mathcal{R}}
\newcommand{\itC}{\mathcal{C}}
\newcommand{\absl}{\left|}
\newcommand{\absr}{\right|}

\newcommand{\libarg}{e_{1}^{\absl j_3 \absr} e_{2}^{\absl j_4 \absr} s_{1}^{\absl j_5 \absr}  s_{2}^{\absl j_6 \absr}}

\documentclass[13pt]{article}
\usepackage{amssymb,amsmath,amstext,amsgen,amsopn,amsxtra,indentfirst}
\numberwithin{equation}{section}
\input psfig

\begin{document}






\newpage

\centerline{\bf \Huge The Dynamics of Two}

\smallskip

\centerline{\bf \Huge Massive Planets}

\smallskip
\smallskip

\centerline{\bf \Huge on Inclined Orbits}

\bigskip
\bigskip

\centerline{\Large by}

\bigskip

\centerline{\Large Dimitri Veras}

\bigskip

\centerline{\Large and}

\bigskip

\centerline{\Large Philip J. Armitage}

\baselineskip=16pt

\begin{abstract}


\Large
\baselineskip=24pt

The significant orbital eccentricities of most giant extrasolar planets may
have their origin in the gravitational dynamics of initially unstable multiple
planet systems.  In this work, we explore the dynamics of two close planets on
inclined orbits through both analytical techniques and extensive numerical
scattering experiments.  We derive a criterion for two equal mass
planets on circular inclined orbits to achieve Hill stability,
and conclude that significant radial migration and eccentricity pumping of
both planets occurs predominantly by 2:1 and 5:3 mean motion resonant
interactions. Using Laplace-Lagrange secular theory, we obtain analytical secular
solutions for the orbital inclinations and longitudes of ascending nodes,
and use those solutions to distinguish between the secular and resonant dynamics
which arise in numerical simulations.  We also illustrate how encounter maps,
typically used to trace the motion of massless particles, may be modified
to reproduce the gross instability seen by the numerical integrations.
Such a correlation suggests promising future use of such maps to model
the dynamics of more coplanar massive planet systems.

\end{abstract}

\bigskip
\bigskip
\bigskip

\noindent{}Keywords:   CELESTIAL MECHANICS, EXTRASOLAR PLANETS,
ORBITS, PLANETARY DYNAMICS, RESONANCES

\newpage


\large
\baselineskip=16pt

\section{Introduction}

Over the last decade, our catalog of planets has increased twelve-fold.
Over 120 extrasolar planets around main sequence stars in over 100 systems
have been discovered, 
\footnote{From the on-line Extrasolar Planets Encyclopedia,
at http://cfa-www.harvard.edu/planets/catalog.html (Schneider 2004) \\ \\as of June 11th, 2004.}
of which at least 10 such systems contain multiple planets.
Only three of the planets around solar-type stars have been discovered by transit
(Konacki {\it et al.} 2003; Konacki {\it et al.} 2004; Bouchy {\it et al.} 2004), and for only 
one other has a transit 
event been detected${}^1$
(Charbonneau {\it et al.} 2000; Henry {\it et al.} 2000).  All other of
these ``Extrasolar Giant Planets'' (EGPs) have been detected by the Doppler radial velocity
technique, which provides information about an exoplanet's semimajor axis and eccentricity, but {\it not} 
its inclination with respect to either other planets or the stellar rotation axis.  The dearth of
data for this one parameter, in addition with the near-coplanarity of the 
major planets in our Solar System, has directed most studies of multiple-exoplanet systems to neglect
any nonzero mutual relative inclination of the planets when discussing the system's past,
present, or future dynamical evolution and resulting stability (an exception is the dynamical study
of the 47 Ursae Majoris system by Laughlin {\it et al.} 2002).

The prospects for two or more planets emerging from a disk on inclined orbits have not yet
been fully explored.  Even in the simplest scenario - in which the planetary disk itself
is flat - excitation of planetary inclination (and eccentricity) can be driven by
mean-motion resonances with the gas disk (e.g. Goldreich and Tremaine 1980), although those
are opposed by dissipative secular effects (Lubow and Ogilvie 2001).  Recently,
Thomas and Lissauer (2003) have used three-dimensional numerical simulations to argue
that non-coplanar planetary systems could be common, provided that damping of {\it eccentricity}
by the disk is relatively weak.  The eccentricity evolution of planets embedded within
disks is, itself, rather uncertain (Papaloizou {\it et al.} 2001; Ogilvie and Lubow 2003;
Goldreich and Sari 2003), and probably depends upon unknown properties of the dynamics of
eccentric disks (Kato 1983; Lyubarskij {\it et al.} 1994; Ogilvie 2001).  In less
standard models - for example those in which the disk itself is warped (Terquem and Bertout 1996),
or the planetary system is initially crowded (Papaloizou and Terquem 2001; Adams and Laughlin 2003;
Nagasawa {\it et al.} 2003) - high relative inclinations between a small number (often two) of
surviving planets are even more probable.  Finally, Yu and Tremaine (2001) conclude that when
an inner planetesimal disk is driven into a star, the runaway multiple planet system may
not be coplanar.

The wide use of the coplanarity assumption has also affected the types of mean motion resonances
analyzed for EGPs.  Resonances may be classified as purely eccentric, purely vertical, or a combination of
the two.  All three types play a crucial role in shaping the dynamical evolution of ring particles,
asteroids, and Kuiper Belt Objects, but for larger bodies like satellites and planets eccentric resonances
dominate; the $4$:$2$ Mimas-Tethys resonance is the {\it only} known purely inclination resonance among
satellites or planets in our Solar System (Peale 1999; Champenois and Vienne 1999a).  However,
Henrard {\it et al.} (1995) and Wisdom (1987) have shown that one must include inclination effects
when attempting to characterize resonances in planetary systems.  Further, most resonance
studies have been conducted under the guise of the restricted three-body problem 
(Peale 1986; Murray and Dermott 1999), or a variation thereof, due to its analytical tractability
and practical application to dynamical situations in our Solar System.  The resonance
between two EGPs cannot be modeled as in the restricted three-body problem because of the unknown
relative inclinations between the EGPs and because neither mass can be treated as negligible.

The coplanarity or small-inclination assumption has affected other types of dynamical analysis as
well.  Gladman (1993) produced pioneering work on establishing the criteria for stability of two-planet
systems, but did not include nonzero planetary inclinations in his analysis.  Traditionally, the
expansion of the disturbing function has been obtained by expanding about inclination values
of $0^{\circ}$ (Kaula 1962; Murray and Dermott 1999).  Although some forms of the disturbing function
expanded about high eccentricity values exist and have been applied to nonplanar problems
(Roig {\it et al.} 1998; Beagu\'{e} and Michtchenko 2003), to our knowledge high inclination expansions
have not yet been derived.  Encounter maps have been used as a technique to trace the orbits of
planetesimal swarms (Namouni {\it et al.} 1996) and outer solar system
objects (Duncan {\it et al.} 1989), but have not yet been used to model highly inclined systems nor systems
with two massive planets.

We address several of the above issues in this work by presenting analytical and numerical
results for two {\it massive} planets on {\it arbitrarily} inclined orbits.  Section 2 provides
a description of the Hill stability limit, and illustrates an extension to the inclined case.  
This extension forms the basis for the $\sim 10^4$ gravitational scattering experiments we
performed, the results of which are presented in Section 3.  In Section 4, we derive compact
expressions for the secular evolution of inclined planets, and compare those to the simulation
results to aid in analyzing the latter.  In Section 5 we illustrate how Namouni {\it et al.'s} (1996)
encounter map may be extended to two massive planets on inclined orbits.  Section 6 discusses
the implications of this work, and Section 7 briefly summarizes the results.

\section{The Hill Stability Limit}

\subsection{Background}

Due to the computational expense and large phase space involved, a numerical study of
even two-planet systems requires a carefully chosen set of initial conditions.
We determined what conditions are most suitable for study by first considering the
types of stability a system may have.  As summarized by Gladman (1993), a system that
is Lagrange stable is one where the planetary orbits
are bound for all time.  A system that is Hill stable is one where the orbits of
planets cannot cross.  Only Hill stability, however, is mathematically tractable.

Marchal and Bozis (1982) applied the mathematical results of Hill stability to two-planet
systems with a central mass much greater than the planetary masses.  Gladman (1993) analyzed
the problem further, providing explicit expressions in terms of orbital elements which
approximate the minimum initial semimajor axis separation of two planets necessary to ensure
Hill stability.  We expand on Gladman's (1993) results by deriving a general expression
for equal-mass planets and an arbitrary initial inclination.

\subsection{Setup}

The Hill stability limit may be considered in terms of the fixed points of the general
three-body problem.  These fixed points are the locations in space that separate
qualitatively different dynamical behaviors of the planets.  Typically, the fixed points
are analyzed for the circular restricted three-body problem, in which the
secondary mass is much greater than the tertiary.  However, not as well studied are the
fixed points of Hill's problem (Hill 1878; Hen\'{o}n and Petit 1986), in which the
{\it secondary and tertiary masses are comparable}.  Hill's problem has wide application
to solar system dynamics, and especially to two-planet systems.

Consider the case of a two-planet system, with planets of mass $m_1$ and $m_2$ orbiting a
massive central object $m_3$ (so that $m_3 \gg m_2$ and $m_3 \gg m_1$).  Then for $m_1 > m_2$,
Marchal and Bozis (1982) showed that the Hill stability limit is given by the solution to: 

\beq
\begin{split}
1 + 3^{\f{4}{3}} \f{m_1 m_2}{m_3^{2/3} {\po m_1 + m_2 \pc}^{\f{4}{3}}} 
-& \f{m_1 m_2 \po 11 m_1 + 7 m_2 \pc}{3 m_3 {\po m_1 + m_2 \pc}^2} + \dotsb 
\\
 &= -\f{2 \po m_1 + m_2 + m_3 \pc}{G^2 \po m_1 m_2 + m_1 m_3 + m_2 m_3 \pc^3} c^2 h,
\label{mb10}
\end{split}
\eeq

\noindent{}where $c$ is the angular momentum and $h$ the total energy of the system.
Note that the left-hand-side (LHS) of this equation is an approximate expansion, which can
be continued if greater accuracy is required.  Since $c$ and $h$ are functions of the
orbital elements, this equation is a (quartic) equation for the separation needed for Hill
stability.

\subsection{The Circular Coplanar Limit}

\subsubsection{Units}

Gladman (1993) provided an analytical solution for the critical separation
in terms of the masses alone.  We adopt his useful notation in this work by setting the
following parameters and defining the following auxiliary variables:  Let $m_3$ be the
``central mass'' so that $m_3 \gg m_1$ and $m_3 \gg m_2$.  Gladman (1993) defined:

\beq
\sum m_{i} = 1, \bbs\bbs \mu_1 \equiv \f{m_1}{m_3}, \bbs\bbs
\mu_2 \equiv \f{m_2}{m_3}, \bbs\bbs {\rm and} \bbs\bbs \alpha \equiv \mu_1 + \mu_2.
\label{g1}
\eeq

\beq
a_1 \equiv 1, \bbs\bbs  a_2 \equiv 1 + \Delta, \bbs\bbs
{\rm and} \bbs\bbs \delta \equiv \sqrt{1 + \Delta}.
\label{g2}
\eeq

\beq
\gamma_i \equiv \sqrt{1 - e_i^{2}}, \bbs\bbs {\rm and} \bbs\bbs G = 1.
\label{g3}
\eeq

The quantity $\Delta$ represents the separation of both planets and will 
be used throughout the rest of this work.  We work in units where the sum
of the masses and the gravitational constant are separately equal to unity.
Since we ignore physical collisions 
between planets, the problem remains scale-free, and
$a_1$ may be scaled straightforwardly to arbitrary values.  For generality, henceforth we quote
semimajor axis in these scaled units (i.e. $a_1 = 1$ initially), and 
{\it times in units of the orbital period at $a=1$}.  Because $m_3$ is considered
to be the central mass, and through the choice of units and approximations, one may
eliminate $m_1$, $m_2$, and $m_3$ in Eq. (\ref{mb10}) in favor of $\mu_1$,
$\mu_2$, and $\alpha$.

\subsubsection{Statement of critical separation}

With Gladman's (1993) variables, the LHS of Eq. (\ref{mb10}) can now be written as:

\beq
1 + 3^{\f{4}{3}} \f{\mu_1 \mu_2}{\alpha^{\f{4}{3}}}
- \f{\mu_1 \mu_2 \po 11 \mu_1 + 7 \mu_2 \pc}{3\alpha^2} + \dotsb
\label{pnu}
\eeq

In route to rewriting the RHS of Eq. (\ref{mb10}) by using the definitions in
Eqs. (\ref{g1})-(\ref{g3}), one must first re-express the 
total energy and angular momentum of the system as such:

\beq
h \approx -\f{\mu_1}{2} - \f{\mu_2}{2 \po 1 + \Delta \pc},
\label{g6}
\eeq

\beq
\begin{split}
c &= \sum_{i=1}^{3} m_i \po \bar{r}_i \times \bar{v}_i \pc \\
  &= G m_1 \sqrt{m_{3}a_1 \po 1 - e_1^{2} \pc} + 
     G m_2 \sqrt{m_{3}a_2 \po 1 - e_2^{2} \pc} + \dotsb \\
&\approx \mu_1 \gamma_1 + \mu_2 \gamma_2 \delta.
\label{g7}
\end{split}
\eeq

Now the RHS of Eq. (\ref{mb10}) can be written as:

\beq
\alpha^{-3} \po \mu_1 + \f{\mu_2}{\delta^2} \pc
{\po \mu_1 \gamma_1 + \mu_2 \gamma_2 \delta \pc}^2.
\label{g8}
\eeq

Equating (\ref{pnu}) and (\ref{g8}) for circular orbits yields
the critical separation of two planets in terms of the planet
masses {\it only} (Gladman 1993):

\beq
\Delta_{crit} = 2 \cdot 3^{\f{1}{6}} {\po \mu_1 + \mu_2 \pc}^{\f{1}{3}} 
+ 2 \left[3^{\f{1}{3}} {\po \mu_1 + \mu_2 \pc}^{\f{2}{3}} -
\f{11\mu_1 + 7\mu_2}{3^{\f{11}{6}}{\po \mu_1 + \mu_2 \pc}^{\f{1}{3}}}     
\right] + \dotsb
\label{g11}
\eeq

If two planets are initially separated by a distances greater than 
$\Delta_{crit}$,
then their orbits cannot cross for all time, and the planets are said to be
Hill stable.  In order to evaluate Eq. (\ref{g11}) for typical giant planet
mass ratios of $\mu = \; \sim 10^{-3}$, the equation may be rewritten in a more practical way, as:

\beq
\f{a_2}{a_1} = 1 + 0.240 \po \f{\mu_1}{10^{-3}} + \f{\mu_2}{10^{-3}} \pc^{\f{1}{3}} + 
\mathcal{O} \po \mu^{\f{2}{3}} \pc
\label{prac}
\eeq

Equation (\ref{prac}) illustrates that for equal mass giant planets, the outer
planet's semimajor axis must be $\sim 30\%$ greater than the inner planet's
to ensure that their orbits never cross.  The critical separation for Jupiter 
and Saturn gives $a_2/a_1 = 1.258$, whereas their actual separation $a_2/a_1 = 1.833$
is much greater and easily satisfies the criterion.

\subsection{The Circular Nonplanar Limit}

In this section we derive a more general form of Eq. (\ref{g11}) that
incorporates an arbitrarily large initial relative inclination between
two equal mass planets.  The angular momentum components ($c_x$, $c_y$, $c_z$)
of a two-body system may be expressed in terms of the angular momentum of
the two bodies ($c_1$, $c_2$) as:

\beq
\begin{split}
c_x &= c_{2_x} = c_2 \sin{I}\sin{\Omega}, \\
c_y &= c_{2_y} = c_2 \sin{I}\cos{\Omega}, \\
c_z &= c_{1_z} + c_{2_z} = c_1 + c_2 \cos{I},  
\label{g12}
\end{split}
\eeq

\noindent{where} $I$ and $\Omega$ are the inclination and longitude of ascending
node of the outer body.  Hence, 

\beq
\vert c \vert^2 = 
\mu_1^{2} \gamma_1^{2} + \mu_2^{2} \delta^2 \gamma_2^{2} +
2 \mu_1 \mu_2 \delta \gamma_1 \gamma_2 \cos{I}
, 
\label{g13}
\eeq

\noindent{} which differs from the square of the planar expression (\ref{g7}) 
only by a factor of $\cos{I}$.  

The energy of the system is only dependent on the masses and relative
semimajor axes of the planets.  Using the new angular momentum, and setting
$\mu_1 = \mu_2 = \mu$, Eq. (\ref{mb10}) yields a quartic equation in $\delta$:

\beq
\f{1}{8} \po 1 + \f{1}{\delta^2} \pc
{\po 1 + \delta^2 + 2\delta\cos{I} \pc}^2 
= 1 + \po\f{3}{2}\pc^{\f{4}{3}} \mu^{\f{2}{3}} - \f{3}{2}\mu + \mathcal{O} \po \mu^{4/3} \pc
.
\label{quartic}
\eeq

\noindent{}The right-hand-side is a slowly converging series in mass ratio
(the ratio of the first two terms containing $\mu$ is just $[2\mu/3]^{1/3}$).
However, the $1$ dominates all terms containing $\mu$ for realistic planet-star
mass ratios, rendering all such terms as minor corrections regardless of their
comparable magnitudes.

Although stability equations such as Eq. (\ref{quartic}) can be solved
numerically for $\delta$, an analytical solution for {\it arbitrary} (not small)
initial relative inclination is possible by using computer algebra.
The exact solution to Eq. (\ref{quartic}) is a series of nested square roots,
with radicands of the
form $\chi + \epsilon(I) + f(I,\mu)$, where $\chi$ is a constant, $\epsilon$ is a
function of the inclination, and $f$ is a double-valued function of
the planetary mass and the inclination.  En route to deriving Eq. (\ref{g11}),
Gladman (1993) encountered radicands of the form $\chi + f(\mu)$, and used
binomial expansions based on the assumption $\chi \gg \mu$.  However,
making the approximation $\chi \gg \mu$  for the radicands with inclination terms
is insufficient, as $\epsilon(I)$ may be comparable to $\chi$ for high inclinations.
Further, one cannot make the approximation $\epsilon(I) \gg \chi$; doing so may
cause the expansion of the radicand to become indeterminate for low inclinations. 
Instead, we use the following expanded form of the binomial approximation:

\beq
{\po \chi + \ep + f \pc}^n = (\chi + \ep)^n {\po 1 + \f{f}{\chi + \ep} \pc}^n
\approx (\chi + \ep)^n {\po 1 + \f{n f}{\chi + \ep} \pc}
.
\label{expbin}
\eeq

Repeated application of Eq. (\ref{expbin}) to the exact solution of Eq. (\ref{quartic})
yields the following result:

\beq
\Delta_{crit} = \ep \lls + \lls
\underbrace{
\eta\sqrt{\po  4 + \f{\cos^2 I}{2} \pc \po \ep + \chi\eta \mu^{\f{2}{3}} \pc}}_
{\mathcal{O}(\mu^{\f{1}{3}})}
\lls + \lls
\underbrace{
\left[\chi\eta \mu^{\f{2}{3}} - 3\eta^2 \mu 
\sqrt{\f{4 + \f{\cos^2 I}{2}}{\ep + \chi\eta \mu^{\f{2}{3}}}} \right]}_
{\mathcal{O}(\mu^{\f{2}{3}})}
 + \dotsb
\label{finali}
\eeq

\centerline{where}

\beq
\ep \lls \equiv \lls 2 + \cos^2 I - \cos I\sqrt{8+\cos^2 I}
,
\label{finali1}
\eeq

\beq
\eta \lls \equiv \lls 1 - \f{\cos I}{\sqrt{8+\cos^2 I}}
,
\label{finali2}
\eeq

\beq
\chi \lls \equiv \lls 3 \cdot 2^{\f{2}{3}} \cdot 3^{\f{1}{3}}
.
\label{finali3}
\eeq

Eqs. (\ref{finali})-(\ref{finali3}) provide the critical separation in terms
of an initial arbitrary relative inclination and planetary mass.  The equations
reduce exactly to Eq. (\ref{g11}) in the limit $I \to 0^{\circ}$, and 
include a term of the order of unity which is not present in Eq.
(\ref{g11}). 

Eq. (\ref{finali}) is in its simplest form because by attempting
to simplify it with additional applications of the binomial
theorem, one would eliminate the $\mu^{1/3}$ term.  Further, because
the planar limit was derived without a small angle approximation, Eq.
(\ref{finali}) allows one to study retrograde planetary motion in addition to prograde
planetary motion.  If both planets are orbiting in opposite directions,
their mutual inclination is $180^{\circ}$.  Fig. \ref{fig1} illustrates 
that $\Delta_{crit}$
rises monotonically with inclination. \marginpar{FIG. 1}  Further, the 
figure illustrates that
the Hill stability limit varies little for any mass ratios
below $10^{-3}$.  In the limit of zero mass and for either perpendicular
or retrograde orbits, the Hill stability limit reduces to compact
rational expressions, as shown in Eq. (\ref{limi}),

\beq
\lim_{\begin{subarray}{1}
      i \to 90^{\circ}  \\
      \mu \to 0        
      \end{subarray} }
      \Delta_{crit} = 2 + 2\sqrt{2} \bs\bs\bs\bs\bs\bs\bs\bs\bs
 \lim_{\begin{subarray}{1}
      i \to 180^{\circ}  \\
      \mu \to 0        
      \end{subarray} }
      \Delta_{crit} = 6 + 4\sqrt{3}
.
\label{limi}
\eeq

Although thus far we have considered arbitrary inclinations, hereafter we
restrict attention to $I \lesssim 40^{\circ}$.  This bound was chosen based
upon the highest inclinations of the known prograde satellites in our solar system.  Systems 
exist with prograde satellites that have inclinations relative
to the ecliptic up to $47^{\circ}$, and are highly unlikely to reside ever in a stable configuration
with inclinations from $55^{\circ} < I < 130^{\circ}$ (Carruba {\it et al.} 2002).

For most extrasolar planetary systems, the mutual inclinations of multiple planets are unknown.
However, it is highly unlikely that multiple massive planets could form with very large
mutual inclinations.  Substantial mutual inclination could subsequently develop - for
example if one of the planets was influenced by a Kozai resonance in a binary
(Holman {\it et al.} 1997) - but even in such cases relative inclinations of tens of 
degrees appear to be improbable.

\section{Simulation Data}

\subsection{Simulation Setup}

\subsubsection{The HNbody integrator}

We performed all our numerical simulations with the HNbody integrator 
(Rauch and Hamilton 2004 in preparation), which
specializes in the integration of systems with a single massive object that comprises the vast majority of
the system mass.  We used the Bulirsch-Stoer integration option with the HNbody accuracy parameter
set to $10^{-12}$ (Rauch and Hamilton 2004 in preparation).  In most cases, energy and angular
momentum errors, expressed by $(h-h_{0})/{h_{0}}$ and $(c_{z}-c_{z_{0}})/c_{z_{0}}$, where
$h_{0}$ is the initial total system energy and $c_{z_{0}}$ is the initial total angular momentum
in the direction perpendicular to the invariable plane, did not exceed $10^{-7}$ over the course
of the run.  In the most pathological cases, energy and z-angular momentum errors were conserved
to within $10^{-4}$.  Angular momenta in the other two directions were typically conserved to
two order of magnitudes better than the z-angular momentum.

\subsubsection{Initial conditions}

\paragraph{Masses, semimajor axes, eccentricities, and inclinations}

The magnitude of the gravitational interaction between two particles results from the masses
and relative positions of the particles.  To facilitate comparison with previous results, we
fixed the planetary masses at $\mu_1 = \mu_2 = 10^{-3}$, where the subscript ``1'' refers to
the inner planet and ``2'' the outer planet.  Ford {\it et al.} (2003) and Ford {\it et al.} (2001)
used these values in their extensive set of simulations, and these values correspond well
to the Sun-Jupiter mass ratio, a benchmark used in the study of many systems with EGPs.  

Besides masses, the separation is the other factor which determines how planets interact.
The parameters $a$, $e$, and $I$ have the largest effect, while the angular variables have
minimal effect.  To isolate the effect of inclination, we begin all simulations with circular orbits
($e_1 = e_2 = 0$).  Further, because any two orbital planes in space have just one mutual
inclination, we set $I_1 = 0^{\circ}$ without any loss of generality.  For the equal mass simulations,
we fixed initial values of $I_2$ in multiples of $5^{\circ}$ ranging from $0^{\circ}$ to $35^{\circ}$.  
We reran the $I_2 = 0^{\circ}$ simulations with $I_2 = 0.001^{\circ}$ in order to avoid
a convergence problem in the code, but will henceforth refer to these simulations as $I_2 = 0^{\circ}$
simulations.  Further, because $I_1 = 0^{\circ}$, the relative initial inclination, denoted
by $I$, will henceforth be equal to $I_2$.

The most interesting dynamical behavior occurs for planetary separations that
most likely give rise to transitory instability followed by quasi-periodic behavior.
Based on the definition of Hill stability, the outer planet must lie between $a_2 = 1$
and $a_2 = 1 + \Delta_{crit}$ in order to make orbit crossing possible.  Therefore, $1 + \Delta_{crit}$
represented the upper bound for $a_2$.  The lower bound, the boundary at which
quasi-periodic behavior largely ceases and at which globally chaotic behavior begins,
cannot be determined analytically, and no previous empirical estimate has
taken into account nonzero inclination.  For the circular, planar case of two equal masses, Wisdom (1980)
found the limit (location) of global chaos to be $\propto \mu^{2/7}$.  Not knowing how this
estimate holds for nonzero inclination, we conducted some preliminary tests, and for ease of
interpretation, chose to establish an inclination-independent lower bound of $a_2 = 1 + \Delta_{crit}/2$.
Having set the range over which $a_2$ varies to be $1 + \Delta_{crit}$ to $1 + \Delta_{crit}/2$, we chose
to decrement $a_2$ in 300 uniform intervals. 

In Fig. \ref{fig2}, \marginpar{FIG 2} we verified that for $I = 0^{\circ}$, 
all stable systems lie beyond $\Delta_{crit}$.
Therefore, this stability criterion, although approximate, appears reasonable for the
coplanar case.  For $I > 0^{\circ}$, we found that most instances of 
instability in this separation
range correspond to resonances.  We explored a fixed range of separations 
($\Delta_{crit}/2 \rightarrow \Delta_{crit}$) to facilitate easy comparison 
between runs.

\paragraph{Angular Variables}

Consider a planet whose orbit is inclined by $I$ with respect to a reference plane.  
The longitude of ascending nodes establish only the orientation of a system
with respect to a reference direction, usually the Earth, and do not affect the dynamics
of the system.  Therefore, we set $\Omega_1 = \Omega_2 = 0$.
Having no reason to suspect a priori that particular values of the initial
mean anomaly would stimulate or inhibit instability, we randomized both $M_1$ and $M_2$ in
the range $[0,2\pi]$ for each one of the sets of 300 simulations we ran.  
We repeated each simulation $4$ times, each corresponding to different (random) mean anomalies.

\paragraph{Time}

Previous studies (Ford {\it et al.} 2001; Adams and Laughlin 2003; Veras and Armitage 2004) have
shown that run times of at least $1 \times 10^6$ are required in order to allow for dynamical settling
of the systems.  We performed a preliminary set of simulations in order to determine
a sufficiently long running time for detection of instability, especially in the region of
interest, the zone outside the global chaos limit.  Figure \ref{fig3} illustrates the timescales
on which systems for four different initial inclinations went unstable. \marginpar{FIG. 3} 
The figure illustrates that
the number of systems which go unstable after $1 \times 10^6$ trails off appreciably.  
Specifically, for $I = (0^{\circ}, 5^{\circ}, 10^{\circ}, 15^{\circ})$, the percent of unstable
systems that went unstable within $1 \times 10^6$ is $(91\%, 85\%, 84\%, 74\%)$.   
This observation,
combined with CPU time and space constraints, led us to choose $2 \times 10^6$ as the length of running
time for all simulations.

\subsubsection{Classifying instability}

We studied only systems in which both planets remained on bound orbits after $2 \times 10^6$; in order to
minimize CPU time, we continuously checked for signatures of instability, and when found
terminated the run.  We classified a system as unstable if at any time the semimajor axis or
eccentricity of either body became negative, or if either body entered the tidal limit of the central
mass.  Weidenenschilling {\it et al.} (1984) provide expressions for the
tidal limit of two spherical particles for a variety of cases.  Both tidal shearing and rotation
have been neglected in this study.  Hence, we chose to adopt the largest possible tidal limit, for
two spherical bodies in synchronous rotation, to our simulations.  This condition may be
written in our variables as:

\beq
L = 2.29 \mu^{-\f{1}{3}} R_{\rm planet}
,
\label{roche}
\eeq

\noindent{where} $L$ is the tidal limit, and $R_{planet}$ is the radius of the planet,
which was taken to be the radius of Jupiter. Using Eq. (\ref{roche}), we terminated
a run when the periapse of either orbit moved within the tidal limit.

\subsection{Numerical Results}

We analyze numerical data in three ways: 1) as a whole, 2) in
sets of 300 runs, 3) for individual runs.  We first consider which systems remained
stable (defined as both planets remaining on bound orbits) over $2 \times 10^6$ and which 
did not.  Figures \ref{fig4}
and \ref{fig5} illustrate the presence or lack of stability for all
simulations performed.  \marginpar{FIG. 4}  \marginpar{FIG. 5}

Figures \ref{fig4} and \ref{fig5} show the stability or instability
of each system as a function the initial semimajor axis separation and
initial relative inclination.  As previously mentioned, four trials were conducted for
each initial set of semimajor axes and relative initial inclination values.  Superimposed
on the plot are the nominal mean motion resonance locations, included to display the
correspondence between these locations
and locations of instability.  Note that the horizontal axes are in terms
of {\it decreasing} initial separation, such that
the leftmost region of the graph represents the critical Hill separation, beyond
which the orbits of the planets cannot cross.

Figures \ref{fig4} and \ref{fig5} illustrate that systems
become unstable either 1) in the global chaos regime  or 2) in the 
non-stochastic regime at first, second and third-order
nominal resonance locations.  
The stable systems of interest are ones where significant dynamical evolution of either or
both planets has occurred.  The greatest measure of this evolution is the maximum semimajor
axis deviation experienced by a planet.  Figures \ref{fig6} and \ref{fig7} isolate systems
where either planet's semimajor axis has deviated by at least $20\%$ from its initial value, and
show that such migrating systems occur either in the region of global chaos, or at major mean motion
resonances.  \marginpar{FIG. 6}  \marginpar{FIG. 7}  We chose $20\%$ based on inspection of the 
data; semimajor axis variation for all systems
was limited to either a few percent, or easily exceeded $20\%$.  We analyze the effective reach of
each of these resonances in Section 4.2, but note here that each system that undergoes significant
radial migration is associated with a resonance.   Further, in both Figs. \ref{fig5} and
\ref{fig7}, for the case of $I = 20^{\circ}$, unstable
systems and migratory stable systems associated with the $2$:$1$ resonance tend to lie on the
starward side of that resonance.  This phenomenon could perhaps be explained by the close
proximity of the $2$:$1$ resonance to the Hill stability limit for $I = 20^{\circ}$, as this
phenomenon doesn't appear to exist for higher values of $I$.
We also point out that Wisdom's (1980) estimation of the global chaos limit, given by the dot-dashed
vertical lines in Figs. \ref{fig4} and \ref{fig6}, corresponds well for inclinations up to
$15^{\circ}$.

We now proceed to analyze individual horizontal rows of data from Figs. \ref{fig4} and
\ref{fig5}.  Figure \ref{fig8} illustrates the main features of most of these strips of data
by displaying the bounded systems (diamonds) which arose from of one set of trials 
performed with relative initial inclinations of $15^{\circ}$ (upper panel),
$25^{\circ}$ (middle panel), and $30^{\circ}$ (lower panel). \marginpar{FIG. 8}  As shown, in most
systems the final value of the outer planet's semimajor axis differs from its
initial value by at most $0.01$.  In the remaining systems, the outer planet may migrate either
inward {\it or} outward.

Consider the upper panel of Figure \ref{fig8}.  The system behavior exhibits a clear demarcation
at an initial outer planet semimajor axis value of about $1.315$.  This demarcation
defines the boundary of global chaos.  When the outer planet begins life within
this boundary, it remains close enough to the inner planet that instability occurs and stochastic
behavior ensues.  When the outer planet lies outside this boundary, it usually remains in a quasi-stable
configuration, but {\it not always}.  Consider the three conspicuous systems for which
$a_2$ $\po \sim 1.41 \pc$.  Although safely outward of the global chaos boundary, these planets
show significant radial migration, up to $5$.  The middle and lower panels illustrate how the
gross system properties change for greater relative initial inclinations.  In the middle panel,
a few systems migrate at around $\sim 1.41$, and roughly $10$ systems
experience migration when  $1.57 < a_2 (initial)< 1.59$.  This same migration regime
presents itself in the lower panel, except with a large clear gap centered about $1.58$.
Note in the middle and lower panels, the systems were sampled only in the non-stochastic
regime.

The radial locations at which planets migrate are those of mean motion resonances.  As the initial
relative inclination increases, these resonances appear to have greater ``reach'', and can more
easily cause a planet to migrate.  This work will focus on identifying the types of resonances
influencing the systems, but will {\it not} delve into the actual interaction at resonance.

Figure \ref{fig9} displays the final eccentricity of the sets of systems 
from the middle panel of Fig. \ref{fig8}.  \marginpar{FIG. 9}  Note that accompanying outward or
inward migration is a large and highly variable change in eccentricity.  Otherwise,
the eccentricity remains close to zero, but is nonzero.  Such small nonzero eccentricities are
unreproducible according to Laplace-Lagrange secular theory, as will be shown,
but may be important in determining the libration widths of eccentric resonances.

\section{Data Analysis}

\subsection{Secular Evolution}

\subsubsection{Introduction}

Understanding the orbital variations of planets away from resonance enables one
to distinguish secular dynamical evolution from resonant dynamical evolution.
Two planets, initially on circular inclined orbits, that settle into quasi-stable orbits
around the central mass vary their orbital elements according to Lagrange's Planetary
Equations, given by Eqs. (\ref{Sub}) for $k = 1,2$ (Brouwer and Clemence 1961), to a good approximation,

\begin{subequations}\label{Sub}
\begin{align}
\f{da_k}{dt} &= \f{2}{n_k a_k} \f{\partial \itR_k}{\partial \lambda_k}, 
\label{Sub1} \\
\f{d\Omega_k}{dt} &= \f{1}{n_k a_{k}^2 \sqrt{1 - e_{k}^2} \sin{I_k}} \f{\partial \itR_k}{\partial I_k}, 
\label{Sub2} \\
\f{dI_k}{dt} &= \f{-\tan{\f{1}{2} I_k}}{n_k a_{k}^2 \sqrt{1 - e_{k}^2}}
\po \f{\partial \itR_k}{\partial \lambda_k} + \f{\partial \itR_k}{\partial \varpi_k} \pc -
\f{1}{n_k a_{k}^2 \sqrt{1 - e_{k}^2} \sin{I_k}} \f{\partial \itR_k}{\partial \Omega_k}
.
\label{Sub3}
\end{align}
\end{subequations}

\noindent{In} Eq. (\ref{Sub}), $\itR_k$ represents the disturbing function for either planet,  $n_k$ the mean
motion, $\lambda_k$ the mean longitude,  $\varpi_k$ the longitude of pericenter, and $\Omega_k$
the longitude of ascending node.  

\subsubsection{Inclination variation}

\paragraph{Laplace-Lagrange theory}

Two planets on circular inclined orbits that remain undisturbed by resonances undergo
sinusoidal inclination variations with the same period
but different amplitudes.  The goal of this section is to explain this behavior
quantitatively, and provide formulas for the time variation
of the inclination of both bodies in terms of {\it only} their masses and initial semimajor
axes.  Laplace-Lagrange secular perturbation theory allows us to derive these analytical formulas
because of the initial conditions chosen and the approximations used; the results agree well
with the numerical simulations.

We apply the secular theory laid out in Murray and Dermott (1999) to our initial conditions
without resorting to numerical computations.   Although this theory relies on several approximations,
which will be delineated, the results obviate more careful but cumbersome calculations.  The first
approximation sets $\sin{I_1} \approx I_1$ and similarly for the outer planet's inclination.
This approximation holds for small angles, but even for angles as large as $30^{\circ}$,
$\po I_1 - \sin{I_1} \pc/\sin{I_1} < 0.05$.  This approximation will
be used in the disturbing function and Lagrange's Equations.  The second approximation is
expanding the averaged disturbing function to {\it only} second order in the secular terms only so that

\beq
\itR_{secular} = -\f{1}{2} \alpha b_{\f{3}{2}}^{\po 1 \pc} 
\po I_{1}^2 + I_{2}^2 \pc + \alpha b_{\f{3}{2}}^{\po 1 \pc} I_{1} I_{2}
\cos{\po \Omega_1 - \Omega_2 \pc}
,
\label{dissec}
\eeq

\beq
\itR_{1} = \f{Gm_2}{a_2} \langle \itR_{secular} \rangle, \bbs\bbs\bbs
\itR_{2} = \f{Gm_1}{a_2} \langle \itR_{secular} \rangle
.
\label{dissecRs}
\eeq

\noindent{Here} $\alpha = a_1/a_2$, the ``$b$'' coefficients are Laplace coefficients,
and the functional form of the other coefficients to second order may be found in Murray and Dermott (1999).
The third approximation is setting $\alpha$ to its initial value for all time.  Inspection
of Eq. (\ref{Sub1}) validates this approximation as that equation shows that for secular
disturbing functions, the semimajor axes remain constant.

With these approximations, Eqs. (\ref{Sub2}) and (\ref{Sub3}) become, for $k = 1,2$,

\begin{subequations}\label{trunlap}
\begin{align}
\f{d\Omega_k}{dt} &= \f{1}{n_k a_{k}^2 I_k} \f{\partial \itR_k}{\partial I_k}, 
\label{trunlap} \\
\f{dI_k}{dt} &=  - \f{1}{n_k a_{k}^2 I_k} \f{\partial \itR}{\partial \Omega_k}.
\end{align}
\label{trunlap}
\end{subequations}

The inclination vectors $p_k = I_k \sin{\Omega_k}$ and $q_k = I_k \cos{\Omega_k}$
allow one to avoid the singularities that Eqs. (\ref{trunlap}) would cause for
low $I_k$ and allow one to redefine the averaged disturbing function as, using $k \ne j$, 

\beq
\langle \itR_k \rangle = n_{k}a_{k}^2 \left[
\f{1}{2} B_{kk} \po p_{k}^2 + q_{k}^2 \pc + B_{kj}
\po p_k p_j + q_k q_j \pc
\right]
,
\label{newR}
\eeq

\noindent{with}

\beq
B_{kj} = C_{kj} \f{m}{4\po m_{\ast} + m \pc} \alpha b_{\f{3}{2}}^{\po 1 \pc} 
.
\label{Bs}
\eeq

Here, we have assumed equal mass planets of mass $m$ and a central mass of $m_{\ast}$.  The
values $C_{kj}$ are given in Murray and Dermott (1999).  The solutions to this problem,
given as inclinations with respect to time, are $I_k(t) = \sqrt{p_{k}^2 + q_{k}^2}$, where

\beq
p_k = \sum_{j=1}^{2} T_{j} I_{kj} \sin{\po f_{j} t + \gamma_{j} \pc}
,
\label{peqs}
\eeq

\beq
q_k = \sum_{j=1}^{2} T_{j} I_{kj} \cos{\po f_{j} t + \gamma_{j} \pc}
.
\label{qeqs}
\eeq

\noindent{The} quantities $T_{j}$ are scaling factors, $\gamma_{j}$ are determined by the initial
conditions, and $f_{j}$ and $I_{kj}$ represent the eigenvalues and eigenvector
components of matrix {\bf B}, with elements given by Eq. (\ref{Bs}).

\paragraph{Application to Inclined Circular Orbits}

The initial conditions for all our simulations force $e_1 = e_2 = \Omega_1 = \Omega_2 = I_1 = 0$.
In each simulation, we set the relative inclination equal to the inclination of the outer planet,
which we denote as $I_0$.  At $t=0$, Eq. (\ref{peqs}) then requires 
$\gamma_1 = \gamma_2 = 0$.  Application of Eq. (\ref{qeqs}), along with our eigenvector
choice of $I_{11} = I_{12} = 1$, gives:

\beq
T_1 = -\f{I_0 I_{12}}{I_{11}I_{22} - I_{12}I_{21}}, \bbs\bbs\bbs\bbs  T_2 = -T_1
.
\label{ra1}
\eeq

Solving for $I_2 (t)$ gives

\beq
I_2 (t) = \sqrt{p_{2}^2 + q_{2}^2} 
= \sqrt{T_{1}^2 I_{21}^2 + T_{2}^2 I_{22}^2 + 2 T_{1}T_{2} I_{21}I_{22}\cos{\po f_1 - f_2 + \gamma_1 - \gamma_2\pc t}}
.
\label{ra3}
\eeq

\noindent{Manipulation} of the two eigenvalues $f_1$ and $f_2$ and use of Eq. (\ref{ra1}) shows that

\beq
I_2 (t) = \f{{I_0}}{1+\alpha^{-\f{1}{2}}} \sqrt{I_{21}^2 + I_{22}^2 - 2 I_{21}I_{22}\cos{\po B_{11} + B_{22} \pc t}}
.
\label{ra22}
\eeq

\noindent{Insertion} of the expressions for the eigenvectors, $I_{kj}$, into Eq. (\ref{ra22}) yields

\beq
I_{2} (t) = \f{1}{1+\sqrt{\alpha}} I_0 \sqrt{1 + \alpha + 2\sqrt{\alpha} \cos{\po \theta t \pc}} 
,
\label{i2t}
\eeq

\noindent{where}

\beq
\theta = -\f{1}{4} b_{\f{3}{2}}^{\po 1 \pc} \f{m}{\sqrt{m_{\ast} + m}}
a_{2}^{-\f{3}{2}} \po \sqrt{\alpha} + \alpha \pc
.
\label{eqtheta}
\eeq

\noindent{A} similar analysis may be performed to derive $I_1 (t)$:

\beq
I_{1} (t) = \f{\sqrt{2}}{1+\sqrt{\alpha}} I_0 \sqrt{1 - \cos{\po \theta t \pc}}
. 
\label{i1t}
\eeq

Equations (\ref{i2t})-(\ref{i1t}) exhibit the behavior seen in our secular
numerical simulations.  The negative sign in Eq. (\ref{eqtheta}) has no apparent physical significance,
but does have a mathematical significance, and will be needed for later computations.  In order
to test the robustness of the approximation provided by Eqs. (\ref{i2t}) and (\ref{i1t}), we use a
numerical simulation that is most likely to give the {\it worst} fit to the equations.  We
obtained such a simulation by 
performing an additional set of $300$ runs with a high initial relative inclination ($35^{\circ}$)
over an initial $a_2$ range of $1.54 - 1.63$, which is almost centered on the $2$:$1$
(lowest possible order) mean motion resonance at $1.5874$.  Therefore, this set of runs samples
systems at an initial $a_2$
increment of $3 \times 10^{-4}$.  The lowest $a_2$ value at which resonantly-induced instability occurs is 
$1.5544$.  We use the system with $a_2 = 1.5541$, the closest system to resonance, to test the goodness
of our analytical equations to secular motions.

Our numerical results exhibit excellent agreement, as seen in Fig. \ref{fig10}.  \marginpar{FIG. 10}
The inclination range of the inner and outer planet,

\beq
0 \bbs \le \bbs I_1(t) \bbs \le \bbs \f{2 I_0}{1+\sqrt{\alpha}} 
,
\label{i1range}
\eeq

\beq
I_0 \f{1 - \sqrt{\alpha}}{1 + \sqrt{\alpha}} \bbs \le \bbs I_2(t) \bbs \le \bbs I_0 
,
\label{i2range}
\eeq

\noindent{correspond} to those seen in the figure.  Note also from
Eq. (\ref{i1range}) that because $\alpha < 1$, the inclination of the inner planet,
which begins its orbit on the reference plane, exceeds the initial
nonzero inclination of the outer planet at the maximum value of $I_1 (t)$.  Further,
Eq. (\ref{i2range}) illustrates that the outer planet, which was inclined to
the reference plane at $t = 0$, never exceeds its original inclination value.
The outer planet oscillates between the reference plane and its initial inclination
value, but never achieves $I_2 = 0^{\circ}$.  Conversely, the amplitude of the inner planet
is greater, as it oscillates between the reference plane and a plane whose inclination
value exceeds $I_0$.

Eq. (\ref{eqtheta}) allows one to estimate the period, $P$, of the oscillations:

\beq
P = \f{8 \pi \sqrt{m + m_{\ast}}}{m b_{\f{3}{2}}^{\po 1 \pc}}
a_{2}^{\f{3}{2}} \po \f{1}{\sqrt{\alpha} + \alpha} \pc
\backsimeq m^{-1}
.
\label{period}
\eeq

All terms in Eq. (\ref{period}) except the masses are of order 1, and because
$m_{\ast} = 1$, the period is mostly dependent on the mass of each planet.  For
Jupiter mass planets, with $m = 0.001m_{\ast}$, the
inclination period of both planets is on the order of $10^3$.

\subsubsection{Node variation}

Having obtained expressions for the inclination of both planets as a function of time,
one may turn around these expressions and find the longitude of ascending nodes
as a function of time.  Taking $f_2$ as the nonzero eigenvalue and equating Eq. (\ref{peqs})
with $p_k = I_k(t)\sin{\Omega_k}$, then Eqs. (\ref{ra1})-(\ref{i2t}), and (\ref{i1t}) yield:

\beq
\Omega_1 (t) = \sin^{-1} \left[
\f{ \sin{\theta t}}{\sqrt{2 \po 1 - \cos{\theta t} \pc}}
\right]
,
\label{node1var}
\eeq

\beq
\Omega_2 (t) = \sin^{-1} \left[
\f{-\sqrt{\alpha} \sin{\theta t}}{\sqrt{1 + \alpha + 2\sqrt{\alpha}\cos{\theta t}}}
\right]
.
\label{node2var}
\eeq

Comparison with the same numerical simulation that is on the verge of instability,
yields Fig. \ref{fig11}, which again shows excellent agreement between the
analytical and numerical results.  \marginpar{FIG. 11}
Note that the variations in $\Omega$ are independent of the initial relative
inclinations, and that $\Omega_1$ becomes indeterminate during each cycle.  However, this
indeterminacy is not a result of the approximations made in Laplace-Lagrange secular theory.
In order to justify this statement, one may use Lagrange's equations in their full generality to
obtain an alternate expression that also exhibits this degeneracy.

The gravitational interaction between two planets, when not in a strong resonance, can be
considered by disturbing functions that incorporate only secular terms.  Hence, with the
inclusion of secular terms up to second order in inclination, we use:

\beq
\langle \itR_1 \rangle =  \langle \itR_1 \rangle_{secular} =  
C_{0} + C_{1} \po s_{1}^2 + s_{2}^2 \pc + C_{2} s_{1} s_{2} \cos{\po\Omega_2 - \Omega_1 \pc}
,
\label{quasi1}
\eeq

\noindent{where} $s_1 = \sin{(I_1/2)}$,  $s_2 = \sin{(I_2/2)}$,  and
$C_{0}$, $C_{1}$, and $C_{2}$  are constants.  Application of
Eq. (\ref{quasi1}) to Eq. (\ref{Sub2}) (setting $e_1 = 0$ without
loss of generality) yields, with $\xi \equiv \Omega_2 - \Omega_1$, 

\beq
\f{d\Omega_1}{dt} = \f{C_{1}}{2 n_1 a_{1}^2} + \f{C_{2}}{4 n_1 a_{1}^2}
\f{s_2 \cos{\xi}}{s_1}
= \f{\absl C_{1} \absr}{2 n_1 a_{1}^2} \left[\f{s_2 \cos{\xi}}
{s_1} - 1 \right]. \label{quasi2}
\eeq

The last equality follows from the relation between $C_1$ and $C_2$ (Murray and Dermott 1999).
The expression for $\dot{\Omega}_2$ is equivalent, except for an interchange of
subscripts.  Eq. (\ref{quasi2}) illustrates that as $I_1$ approaches zero, $\dot{\Omega}_1$ becomes
indeterminate.  Further, Eq. (\ref{quasi2}) illustrates that the condition for the longitude of nodes to
increase may be expressed as:

\beq
\begin{split}
\dot{\Omega}_1 > 0 \bs\bs {\rm when} \bs\bs &\f{s_2 \cos{\xi}}{s_1} > 1, \\
\dot{\Omega}_2 > 0 \bs\bs {\rm when} \bs\bs &\f{s_1 \cos{\xi}}{s_2} > 1.
\label{quasi3}
\end{split}
\eeq

Further, differentiating Eq. (\ref{node1var}) gives

\beq
\dot{\Omega}_1(t) = -\f{\theta}{2},
\label{om1var}
\eeq

\noindent{a} constant.  Eqs. (\ref{quasi3}) and (\ref{om1var}) provide a mathematical
basis for the time rate of change of $\Omega_1$ seen in Fig. \ref{fig11}.  Differentiation
of Eq. (\ref{node1var}) yields a somewhat, less cleaner result (than Eq. \ref{om1var}) that
contains $\alpha$ to take into account the curvature seen in Fig. \ref{fig11}.  Further, integrating
Eq. (\ref{om1var}) directly illustrates that $\Omega_1$
is a linear function of time, but leaves an undetermined constant that is
absent from the exact Eq. (\ref{node1var}).

\subsubsection{Eccentricity variation}

Unlike for a planar eccentric system, in which $I = 0$ for all time,
for a nonplanar circular system, $e \ne 0$ for all time.  Hence, both planets develop a
nonzero eccentricity, which can play a crucial role in determining the type of resonance
acting in a system.  However, second-order Laplace-Lagrange secular
theory cannot reproduce these nonzero eccentricities, instead predicting that $e = 0$ for
all time.  This result, which may be presupposed based on the decoupling of eccentricity
and inclination in second-order Laplace-Lagrange theory, is often made heuristically
using degeneracy considerations.  In this section, we confirm this result rigorously, linking
the eccentricity and inclination through the Lagrange equation for ${d \Omega}/{d t}$.
Having developed the formalism to derive analytical formulas for the circular inclined case,
we then briefly show how to obtain similar equations for the coplanar eccentric case,
and the complications that result from doing so.

\paragraph{Forever circular orbits}

The Lagrange Equation given by Eq. (\ref{Sub2}) is an exact formula, derived without approximation.
We will use this formula to solve for the eccentricity as a function of time,
as we now know all the other quantities in the formula from secular theory.  These ``other quantities''
are in good agreement with the numerical results.  The secular part of the disturbing
function expanded to second order in {\it both} eccentricities and inclinations 
may be written as (Murray and Dermott 1999):

\beq
\itR_{D}^{\rm secular} = K_1 \po e_{1}^2 + e_{2}^2 \pc 
- \f{1}{2} \alpha b_{\f{3}{2}}^{\po 1 \pc} \po s_{1}^2 + s_{2}^2 \pc
+ K_2 e_1 e_2 \cos{\po \varpi_1 - \varpi_2 \pc} + \alpha b_{\f{3}{2}}^{\po 1 \pc} 
s_1 s_2  \cos{\po \Omega_1 - \Omega_2 \pc}
\label{new2R}
\eeq

\noindent{where} $K_1$ and $K_2$ are constants that will be irrelevant to the subsequent
analysis.  Equation (\ref{new2R}) may be used in the expressions for both disturbing functions
of the problem, such that (Murray and Dermott 1999):

\beq
\itR_1 = \f{Gm_2}{a_1} \alpha \itR_{D}^{\rm secular}  \bbs\bbs\bbs
\itR_2 = \f{Gm_1}{a_1} \alpha \itR_{D}^{\rm secular}
.
\label{totalR}
\eeq

Taking the partial derivative appropriate to Eq. (\ref{Sub2}) gives:

\beq
\f{\partial \itR_1}{\partial I_1} =  \f{G m_2 \alpha^2 b_{3/2}^{\po 1 \pc}}
{2 a_1} \cos{\po\f{1}{2} I_1\pc} \left[s_2 \cos{\xi}  - s_1
\right]
.
\label{parR}
\eeq

Inserting Eq. (\ref{parR}) into Eq. (\ref{Sub2}) and solving for $e_1$ gives:

\beq
e_1 (t) = \sqrt{1 - 
\po{ \f{
G m_2 \alpha^2 b_{3/2}^{\po 1 \pc} \left[
s_2(t) \cos{\xi(t)} - s_1(t)
\right]
}
{
4 a_{1}^{\f{3}{2}} \sqrt{m_{\ast} + m} \dot{\Omega}_1(t) s_1(t)
}
}\pc^2
}
.
\label{thee1}
\eeq

Similarly, 

\beq
e_2 (t) = \sqrt{1 - 
\po{ \f{
G m_1 \alpha b_{3/2}^{\po 1 \pc} \left[
s_1(t) \cos{\xi(t)} - s_2(t)
\right]
}
{
4 a_{2}^{\f{3}{2}} \sqrt{m_{\ast} + m} \dot{\Omega}_2(t) s_2(t)
}
}\pc^2
}
.
\label{thee2}
\eeq

Equations (\ref{thee1}) and (\ref{thee2}) illustrate how both planets' eccentricities
vary through time due to secular effects alone.  Knowing an analytic expression for the eccentricities
at a particular time, one may now proceed with a secular analysis similar to the one performed for
the planets' inclination.  For a disturbing function that includes eccentricity
terms $A_{jk}$, and for scaled eccentricities $h_{k} = e_{k} \sin{\varpi_{k}}$ and
$\kappa_{k} = e_{k} \cos{\varpi_{k}}$, one may write down a set of equations similar
to Eqs. (\ref{Bs})-(\ref{qeqs}) (Murray and Dermott 1999),

\beq
A_{jj} = C_{jj} \f{m}{4\po m_{\ast} + m \pc} \alpha b_{\f{3}{2}}^{\po 1 \pc} 
,
\label{As1}
\eeq

\beq
A_{kj} = C_{kj} \f{m}{4\po m_{\ast} + m \pc} \alpha b_{\f{3}{2}}^{\po 2 \pc} 
,
\label{As2}
\eeq

\beq
h_k = \sum_{j=1}^{2} S_{j} e_{kj} \sin{\po g_{j} t + \beta_{j} \pc}
,
\label{heqs}
\eeq

\beq
\kappa_k = \sum_{j=1}^{2} S_{j} e_{kj} \cos{\po g_{j} t + \beta_{j} \pc}
.
\label{keqs}
\eeq

Note that because of the lack of ``mutual eccentricity'' (like mutual inclination), the
coefficients $A_{jk}$ lack the symmetry of the $B_{jk}$ terms, which all had the
same Laplace ``$b$'' constants.  This prevents significant simplification of the
roots in the eigenvalue ($g_{k}$) and eigenvector ($e_{jk}$) expressions.  Further,
the phase angles $\beta$ now are nonzero because of the temporal conditions used.

Earlier we could not apply Laplace-Lagrange secular theory to the eccentricities,
because initially $e_1 = e_2 = 0$, and hence the secular equations were degenerate
at the only known temporal condition.  Now, however, with Eqs. (\ref{thee1}) and (\ref{thee2}),
we may obtain another time condition (not necessarily an {\it initial} condition).
Denote by $t_{0}$ the time at which the inclinations of both planets are equal, a condition
which occurs twice in every inclination cycle.  Equations (\ref{i2t}) and (\ref{i1t}) then
show that $t_{0}$ satisfies

\beq
\cos{\po\theta t_{0}\pc} = \f{1-\sqrt{\alpha}}{2}
.
\label{i1eqi2}
\eeq

Evaluating $e_{1} (t_0)$ using Eqs. (\ref{i2t})-(\ref{i1t}), (\ref{node1var})-(\ref{node2var}), 
(\ref{om1var}), (\ref{thee1}), and (\ref{i1eqi2}) yields, after much algebra,

\beq
e_1 (t_0) = 0
.\label{itszero}
\eeq

One may also show that $e_2 (t_0) = 0$.  Hence, because the eccentricities are zero at
both $t = 0$ and $t = t_0$, one may use Eqs. (\ref{heqs}) and (\ref{keqs}) to illustrate
that $\beta_1 = \beta_2$, and hence $g_1 = g_2$.  However, as shown by 
Eqs. (\ref{As1})-(\ref{As2}), the two eigenvalues of matrix {\bf A} do {\it not} satisfy
$g_1 = g_2$.  This contradiction forces both eccentricities to be zero for all time,
an unphysical result.  Hence, secular theory cannot predict the eccentricity evolution
of two planets on initially circular orbits, even if those planets do not ever
drift into a strong mean motion resonance.

\paragraph{Planar eccentric case}

Although not directly relevant to the results of this work, an expansion of the previous results
to the planar eccentric case could perhaps prove useful for future studies.  Just as previously
we set the eccentricities and nodes to zero, here we set $I_1 = I_2 = \varpi_1 = \varpi_2 = 0$.
However, unlike inclination, eccentricity is not defined to be a relative quantity.  If we
set $e_1(t=0) = 0$ and $e_2(t=0) = e_{20}$, then

\beq
S_1 = -\f{e_{20} e_{12}}{e_{11}e_{22} - e_{12}e_{21}}, \bbs\bbs\bbs\bbs  S_2 = -S_1 \f{e_{11}}{e_{12}}
.
\label{pe1}
\eeq

Otherwise, if we set $e_2(t=0) = 0$ and $e_1(t=0) = e_{10}$, then

\beq
S_1 = \f{e_{10} e_{22}}{e_{11}e_{22} - e_{12}e_{21}}, \bbs\bbs\bbs\bbs  S_2 = -S_1 \f{e_{21}}{e_{22}}
.
\label{pe2}
\eeq

Solving for $e_k (t)$ gives

\beq
e_k (t) = \sqrt{h_{k}^2 + \kappa_{k}^2} 
= \sqrt{S_{1}^2 e_{k1}^2 + S_{2}^2 e_{k2}^2 + 2 S_{1}S_{2} e_{k1}e_{k2}\cos{\po g_1 - g_2 + \beta_1 - \beta_2\pc t}}
.
\label{ra3}
\eeq

Regardless of the choice of a nonzero $e_{10}$ or $e_{20}$, $\beta_1 = \beta_2$ from 
Eq. (\ref{heqs}).  Hence, one may obtain expressions with the same functional form as 
Eqs. (\ref{i2t}) and (\ref{i1t}) for eccentricity.  Direct computation of the eigenvectors
$g_1$ and $g_2$ (Murray and Dermott 1999, p. 318) shows that 

\beq
\theta \equiv g_1 - g_2 = -\f{1}{4} \mu_{1}^{\f{3}{2}} a_{1}^{-\f{3}{2}}
\alpha^2 b_{\f{3}{2}}^{(1)} 
\sqrt{
\po \f{m_2}{m_1} \pc^2
- 2\f{m_2}{m_1}
\left[
1 - 2 \po \f{b_{3/2}^{(2)}}{b_{3/2}^{(1)}} \pc^2 
\right] \alpha^{\f{1}{2}}
+ \alpha
}
\label{ecctheta}
\eeq

The asymmetry produced by computing eccentricity instead of inclination is contained entirely
within the ratio of Laplace coefficients in Eq. (\ref{ecctheta}); when this term becomes
unity, and in the limit of equal mass planets, Eq. (\ref{ecctheta}) reduces to 
Eq. (\ref{eqtheta}) exactly.  Thus, the deviation in the period of oscillations in the inclined
circular case to that from the planar eccentric case goes as, when $\alpha \ll 1$:

\beq
1 - \f{b_{3/2}^{(2)}}{b_{3/2}^{(1)}} \approx 1 - \f{5}{4} \alpha
\label{devcases}
\eeq

\noindent{where} polynomial expansions for the Laplace coefficients in $\alpha$ can be
found in Murray and Dermott (1999).

\subsection{Libration Widths}

By inspection of Figs. \ref{fig4}-\ref{fig7}, one might conclude that all unstable or
significantly evolving systems that initially lay outside of the global region of chaos are
associated with a resonance.  This section will attempt to support this hypothesis by classifying
the likely resonances that are interacting, and quantifying the effective ``reach'' of such resonances.

Two bodies are said to be ``captured'' into a resonance when one or both of the bodies
drifts closer to the other until a commensurability is attained.
If the planets are moving away from each other, resonant
capture cannot occur (Sinclair 1972; Yu and Tremaine 2001).  However, resonant capture 
is not assured if two planets approach each other.  The perturbations
on two planets in quasi-stable orbits alone does produce instances where the planets
are momentarily drifting toward one another, but for our simulations, this drift was not
sufficient to cause any resonant capture.

Just because a planet's initial semimajor axis lies close to a strong resonance does not mean the planet
will achieve a state of near-exact resonance with that particular commensurability.  Further, a planet
may undergo a series of resonant encounters, each on a different timescale, so as to obliterate any
record of its previous locations.  These considerations should be remembered when comparing
the initial and final semimajor axes of the planets studied here.  Further, a planet has zero
probability of achieving a particular exact resonance, but a finite probability of drifting
inside a relevant interval centered on an exact resonance.  This interval is most typically
classified as the ``libration width'', and is relevant in the sense that a planet drifting
inside may suffer a significant radial excursion if not captured by the resonance.

The gaps in the strings of black diamonds in Fig. \ref{fig8} appear to increase with
the initial relative inclination.  These gaps visually may be thought of as libration widths.  In order
to affirm and quantify this trend of increasing libration
gap width with inclination, we now proceed to derive an analytical estimate of the gap width.
Murray and Dermott (1999) provide such an estimate for the eccentric, planar case with one negligible
mass; we consider the nonplanar, non-circular case with no assumptions about the masses.

Suppose one wishes to find the libration width of a resonance whose resonant argument is
$\phi = j_1 \lambda_1 + j_2 \lambda_2 + j_3 \varpi_1 + j_4 \varpi_2 + j_5 \Omega_1 + j_6 \Omega_2$.
One of the disturbing functions may then be expressed as:

\beq
\langle \itR_1 \rangle = \f{Gm_2}{a_2} 
\lbrack \langle \itR_1 \rangle_{secular} +  \itC_1 \libarg \cos{\phi} \rbrack
.
\label{lag3}
\eeq

A complete description of the nontrivial computation of $\itC_1$ may be found
in Murray and Dermott (1999).  Application of Eq. (\ref{Sub1}) yields

\beq
\f{da_1}{dt} = \f{2}{n_1 a_1} \f{\partial \langle \itR_1 \rangle}{\partial \lambda_1} = 
-\f{2 G m_2 j_1}{n_1 a_1 a_2} \itC_1 \libarg \sin{\phi}
.
\label{lib1}
\eeq

\noindent{which may be rewritten as}

\beq
\dot{n}_1 =  \f{3 G m_2 j_1}{a_{1}^2 a_2} \itC_1 \libarg \sin{\phi}
.
\label{lib2}
\eeq

Similarly, the other disturbing function for the problem can be used to write:

\beq
\dot{n}_2 =  \f{3 G m_1 j_2}{a_{2}^3} \itC_2 \libarg \sin{\phi}
.
\label{lib3}
\eeq

Ignoring the second derivatives of angular variables, one obtains:

\beq
\ddot{\phi} = j_1 \dot{n}_1 + j_2 \dot{n}_2 = 
   \underbrace{\libarg \po \f{3 G m_2 j_{1}^2}{a_{1}^2 a_2} \itC_1 + \f{3 G m_1 j_{2}^2}{a_{2}^3} \itC_2 \pc}_{\omega_{0}^2}  
\sin{\phi}
.
\label{lib6}
\eeq

Equation (\ref{lib6}) is the equation of a pendulum, where the angular frequency
$\omega_{0}^2$ may be represented as the coefficient of $\sin{\phi}$. Comparing the maximum to minimum
energy of this pendulum, and relating the result to Eq. (\ref{lib3}) yields:

\beq
dn_2 = \pm \f{\omega_0}{j_2} \sin{\po \f{1}{2} \phi \pc} d\phi 
.
\label{lib7}
\eeq

Integrating Eq. (\ref{lib7}) and converting to $a_2$ gives the libration
width, $\delta a_{2,lib}$:

\beq
\delta a_{2,lib} = \pm \f{4 a_2 \omega_0}{3 n_2 j_2} = \pm \f{4 a_2}{3 n_2 j_2}
\sqrt{\libarg \po \f{3 G m_2 j_{1}^2}{a_{1}^2 a_2} \itC_1 + \f{3 G m_1 j_{2}^2}{a_{2}^3} \itC_2 \pc}
\propto e_{1}^{\f{\absl j_3 \absr}{2}} e_{2}^{\f{\absl j_4 \absr}{2}} s_{1}^{\f{\absl j_5 \absr}{2}} 
s_{2}^{\f{\absl j_6 \absr}{2}}
.
\label{lib8}
\eeq

Equation (\ref{lib8}) compares favorably with a similar expression derived by Champenois and Vienne
(1999b) to study the Mimas-Tethys $4$:$2$ resonance.  One observes that $\delta a_{2,lib}$ monotonically
increases with an increase in
the eccentricity or inclination of either body, and for two equally massive planets is $\sim \sqrt{2}$
greater than that of the restricted three-body case.  Further, the d'Alembert relations require
that the order of a resonance equals $\absl j_3 + j_4 + j_5 + j_6\absr$, and that the
quantity $\absl j_5 + j_6\absr$ be even.  Hence, any first and third-order resonances {\it must}
include a contribution from one of the planetary eccentricities.  Therefore, the inability
of secular theory to reproduce the observed eccentricity evolution unfortunately shrouds
one's ability to determine which resonances are acting.  Further, because the eccentricities
can reach in excess of $0.01$, their contribution to the libration width often rivals the
contribution of the inclinations, preventing one from neglecting the eccentricities in
resonance calculations.  For second-order and higher-order resonances, with planetary eccentricities
of $0.03$, an initial relative inclination of $10^{\circ}$ produces libration widths
that are comparable to those produced by eccentricities.  Any higher initial inclination
will cause vertical resonances to dominate.

Being functions of eccentricity and inclination, libration widths should also be a function
of time.  In Fig. \ref{fig12} we provide a sample of the magnitudes of the libration
widths for two independent-of-eccentricity $5$:$3$ vertical resonances (the ``mixed''
resonance with $j_5 = j_6 = -1$, and the ``unmixed'' resonance with either $j_5 = -2$ or $j_6 = -2$),
using Eqs. (\ref{i2t}) and (\ref{i1t}) to model the time evolution.  \marginpar{FIG. 12} As shown in Fig.
\ref{fig12}, an unmixed vertical resonance is most likely to take effect in the middle of a 
dynamical inclination period, and a mixed vertical resonance is most likely to take effect at
the beginning (or end) of a dynamical inclination period.

The large phase space covered by our simulations necessitated
an undersampling of the data, preventing us from exploring the detailed interactions at resonance.
Figures \ref{fig4}-\ref{fig7} illustrate that the $2$:$1$ resonance location
affects a relatively large number of systems.  The $2$:$1$ resonance is a first-order resonance,
implying that planetary eccentricity, {\it only}, determines the dynamical evolution of
systems residing in that commensurability.  However, the 2nd-order $4$:$2$ resonance,
3rd-order $6$:$3$ resonance, or such higher order resonances may be acting instead,
and thus inclination may play a role as well.  Only detailed, small-timestep calculations
of custom-models of such resonances can further pinpoint the dynamics induced there.

Table 1 documents the effect that every first through third-order resonance which
overlapped with the phase space studied has on all systems that exhibited significant
dynamical evolution (where either semimajor axis varied by over 20\%). \marginpar{TABLE 1} Of the $9600$
systems modeled, $87$ stand out because they:  1) remained stable over $2$ Myr, 2)
began outside the region of global chaos, 3) and have shown significant radial migration 
of at least one planet.  Table 1 illustrates that $85$ of these $87$ systems
($98\%$) had initial configurations where both planets lay within two libration widths
of a 1st, 2nd, or 3rd-order resonance.  In the table, columns labeled $\pm 1w$ and
$\pm 2w$ list the number of systems that initially lay within one and two libration
widths respectively of the resonance location in question.  The libration widths used
were the maximum widths (computed from Eq. \ref{lib8}) attained over the secular
evolution of the systems, which were modeled using Eqs. (\ref{i2t}) and (\ref{i1t}) for the
inclinations and with fixed orbital eccentricities of $0.03$ (an upper bound on the eccentricities
seen in the simulations during secular evolution) for each planet.  The rightmost column
displays the percent of systems exhibiting significant radial migration
that initially lay within two libration widths of any 1st-order thru 3rd-order resonance
for given relative initial inclinations, $I$.  For $I = 20^{\circ}$ and $I = 25^{\circ}$,
some systems resided within twice the libration widths of both the $8$:$5$ and $5$:$3$ resonances, 
or both the $5$:$3$ and $7$:$4$ resonances.  The bottom row shows that $84\%$ of systems that
fulfilled the three criteria above were affected by the $2$:$1$ or $5$:$3$ resonances.
The two systems not ``picked out'' by the conditions stated above initially lay within $0.01$ 
of the fourth-order $7$:$3$ resonance, located at $1.759$.  This location represents the largest
outer planet semimajor axis for which two planets on inclined, circular orbits exhibit radial migration,
as Table 1 illustrates that despite residing within the region that allows crossing orbits,
the stronger but further out $5$:$2$ and $3$:$1$ resonances do not induce such behavior.

\section{Mapping Results}

\subsection{Background}

We now investigate the extent to which encounter maps can be manipulated to reproduce the
behavior seen in the integrations.  Encounter maps
represent useful tools for describing the geometry of successive conjunctions between
a test particle and a massive perturber, one which is orbiting a more massive central 
body.  These maps also have been applied to the orbits of planetesimal
swarms (Namouni {\it et al.} 1996, henceforth referred to
as NLTP) and outer solar system objects (Duncan {\it et al.} 1989).  The primary
advantage of maps over numerical integrations is the $10$ to $100$-fold decrease
in computation time.  Although these maps traditionally model systems with a massless
test particle, we show that they may be used to model systems with two equal mass planets, and
compare the results of these maps with our numerical integration results.

Duncan {\it et. al} (1989) derived an encounter map for the case where
the inclinations of the test particle orbit and perturber orbit are zero,
the initial eccentricity of the test particle is small, and the initial eccentricity of the
perturber is carried out to first order.  Their first-order map used the small
eccentricity approximation given by Julian and Toomre (1966).  NLTP
derives a more general map that allows for initial eccentricities
and inclinations (which will be henceforth denoted by lowercase ``$i$'', in order to maintain
consistency with NLTP's variables) of the perturber to be expanded to arbitrary order about
$e = 0$ and $i = 0$.  Despite this expansion about zero inclination, we show that
NLTP's map, to second-order only, can faithfully reproduce the qualitative gross
features of the relatively high-inclination systems we have studied.  With this map,
we then extrapolate to semimajor axis regions of interest that we did not
explore with the HNbody code due to CPU constraints.

\subsection{NLTP's Map Generating Algorithm}

NLTP derives a planar, second-order encounter map, by using the variables
and equations we summarize in Table \ref{vars}. \marginpar{TABLE 2} In order to maintain consistency
with the rest of this study, we define the quantities they use as the mass $m$,
semimajor axis $a$, eccentricity $e$, inclination $i$, longitude of perihelion $\omega$,
and longitude of ascending node $\Omega$.  They further define the mean longitude $\lambda$,
true longitude $\theta$, Hill parameter $\epsilon$, and action angle variables $I,J,$ and $K$.
The Hill parameter scales $a$, $e$, and $i$ into Hill units, which enable NLTP to derive
the desired mapping algorithm. 

The quantity NLTP use to perform expansions in eccentricity and inclination out to arbitrary
order is an effective potential, defined by $W$.  This potential is derived from Hill's equations, where
Hill's equations are obtained from the following approximations:

$$m_2,m_3 \ll m_1, \hspace{1.0cm} e_2,e_3 \ll 1,$$

$$i_2,i_3 \ll 1, \hspace{1.0cm} |a_2 - a_0| \ll a_0, \hspace{1.0cm} |a_3 - a_0| \ll a_0.$$

$W$ takes the form of a Fourier series, where, with $j = \sqrt{-1}$, $n\in \mathcal{Z}$,

\beq
W = \sum_{n=-\infty}^{\infty} W_n e^{jn(\lambda-\omega)}
,
\label{W}
\eeq

\beq
\begin{split}
W_{n} = \thickspace &\frac{-2}{a\pi} \int_{-\pi}^{\pi}
e^{-jn(\frac{4}{3}e_{\ast}\sin{(t-\omega)} + t-\omega)}, \\
&K_{0} \left(
\frac{2n}{3}
{\left({(e_{\ast}\cos(t-\omega) + 1)}^2 + i_{\ast}^{2}\cos^2(t-\Omega)\right)}^{\frac{1}{2}}
\right) 
\,d(t-\omega)
.
\end{split}
\label{Wexact}
\eeq

Here $K_0$ is the zeroth-order modified Bessel function of the second kind.
Equation (\ref{Wexact}) gives the exact form of $W_n$.  For details involved
in the derivation of this ``interaction potential'', we refer the interested
reader to NLTP.  \footnote{One should be aware of two typos in NLTP:  the cosine
and sine in their Eqs. (37) and (38) should be interchanged.}  In order to
use different orders of approximation, NLTP expand this equation in the following form: 

\beq
W_{n} = \frac{1}{a}  \sum_{\substack{|n|\le p+2q \le M\\ parity(n)=parity(p)}} 
W_{n}^{p,q}e_{\ast}^{p}i_{\ast}
^{2q} + \hdots
\label{Wexpansion}
\eeq

The quantity $W_{n}^{p,q}$ is expressed in terms of Bessel functions and the difference, ($\Omega-\omega$).
The terms $W_{n}^{p,q}$ are generally a function of $(\Omega-\omega)$ only,
but are constants when $q = 0$.   For $0 < q \le 4$, $W_{n}^{p,q}$ is a function
of $\Omega$ only.  Values for many of these terms are given in Appendix B of NLTP, and
are not be reproduced here.  $M$ represents the order of the approximation of the desired map,
while $p$ and $q$ are integers that do not exceed $M$.  Hence, the second-order planar map
that NLTP derive corresponds to $M = 2,$ $q = 0$.

The effective potential $W$ is used to derive the encounter map
from the following equations:

\beq
I_{n+1} = I_n + \frac{\partial}{\partial \phi}
W(\omega_n,\Omega_n,\lambda_n,I_{n+1},J_{n+1},K_{n+1})
,
\label{Iexact}
\eeq

\beq
J_{n+1} = J_n + \frac{\partial}{\partial \psi}
W(\omega_n,\Omega_n,\lambda_n,I_{n+1},J_{n+1},K_{n+1})
,
\label{Jexact}
\eeq

\beq
K_{n+1} = I_{n+1} + J_{n+1} - I_n - J_n + K_n
,
\label{Kexact}
\eeq

\beq
\omega_{n+1} = \omega_n - \frac{\partial}{\partial I}
W(\omega_n,\Omega_n,\lambda_n,I_{n+1},J_{n+1},K_{n+1})
,
\label{omegaexact}
\eeq

\beq
\Omega_{n+1} = \Omega_n - \frac{\partial}{\partial J}
W(\omega_n,\Omega_n,\lambda_n,I_{n+1},J_{n+1},K_{n+1})
,
\label{Omegaexact}
\eeq

\beq
\lambda_{n+1} = \lambda_n - \frac{\partial}{\partial K}
W(\omega_n,\Omega_n,\lambda_n,I_{n+1},J_{n+1},K_{n+1}) +
\frac{4\pi}{3\epsilon}\sqrt{\frac{3}{8K_{n+1}}}
.
\label{lambdaexact}
\eeq

Note that $I_{n+1}$, $J_{n+1}$ and $K_{n+1}$ appear implicitly in the above equations.
As a result, solving for the complete map analytically in closed
form is impossible for all but the lowest order approximations.
Therefore, NLTP ``keep only the free motion contribution to the
longitude'' by setting $a_r = a_2$ in Eqs. (\ref{Iexact})-(\ref{lambdaexact}).
Additionally, they dispose of the differential term in Eq. (\ref{lambdaexact})
because they claim that this approximation holds when the bodies, ``do not suffer
large excursions in space.''  Because we are using NLTP's algorithm only to aid in
our effort of quantifying what initial conditions produce instabilities, and not
to explore the evolution of any one particular system, we zero out the differential term
in Eq. (\ref{lambdaexact}) as well.

One can see that from Eq. (\ref{Wexpansion}) that when $W_n$ is expanded, many
terms with a linear dependence on eccentricity will result.  These terms, from the definition
of the action-angle variable $I$, are proportional to $\sqrt{I}$.  Then from Eq.
(\ref{omegaexact}) of the map, one sees that the derivative of $W$ is taken with respect
to $I$, resulting in singularities for the aforementioned terms.
In order to circumvent these singular terms, NLTP first splits the potential, $W$,
identifying the portion that contains the singular terms as $W_{sing}$.  NLTP then 
iterates $W_{sing}$ with the symmetry-preserving variables,
$h$ and $k$, to create an intermediate step between $n$ and $n+1$ that removes any degeneracy.
With primed variables representing the intermediate step, the resulting equations read:

\beq
h' = h_{n} + \frac{\partial}{\partial k} 
W_{sing}(k_{n}, \Omega_{n}, \lambda_{n}, h', J', K')
,
\label{hpr}
\eeq

\beq
k' = k_{n} - \frac{\partial}{\partial h} 
W_{sing}(k_{n}, \Omega_{n}, \lambda_{n}, h', J', K')
,
\label{kpr}
\eeq

\beq
I' = \f{h' + k'}{2}
,
\label{Ipr}
\eeq

\beq
J' = J_{n} + \frac{\partial}{\partial \Omega} 
W_{sing}(k_{n}, \Omega_{n}, \lambda_{n}, h', J', K')
,
\label{Jpr}
\eeq

\beq
K' = \f{h'{}^2 + k'{}^2 - h_{n}^2 - k_{n}^2}{2} + J' - J_{n} + K_{n}
,
\label{Kpr}
\eeq

\beq
\omega' = \arctan{\f{k'}{h'}}
,
\label{ompr}
\eeq

\beq
\Omega' = \Omega_{n} - \frac{\partial}{\partial J} 
W_{sing}(k_{n}, \Omega_{n}, \lambda_{n}, h', J', K')
,
\label{Ompr}
\eeq

\beq
\lambda' = \lambda_{n} - \frac{\partial}{\partial K} 
W_{sing}(k_{n}, \Omega_{n}, \lambda_{n}, h', J', K')
.
\label{lampr}
\eeq

Similarly to Eqs. (\ref{Iexact})-(\ref{Kexact}), Eqs. (\ref{hpr})-(\ref{lampr})
are implicit functions, and thus inhibit one's ability
to write down explicit analytic expressions for higher order maps.  Equations
(\ref{Iexact})-(\ref{lampr}) represent the algorithm one may use to derive NLTP's 
encounter map.  The number of terms kept in the
Taylor expansion of $W$ (Eq. \ref{Wexpansion}) determines the accuracy
of the map.  We now use this algorithm to derive a second-order
spatial map that is applicable to the systems integrated in this work.

\subsection{The Nonplanar Second-Order Map}

NLTP use planar maps of up to tenth order, and provide the full expression
for the second-order eccentric planar map, but do not study nonplanar maps.
In this section, we use NLTP's general mapping algorithm to create a second-order
eccentric nonplanar map, which correctly reduces to NLTP's second-order eccentric
planar map in the limiting case of zero inclination.  Then we show that with a
modification, the second-order eccentric nonplanar encounter map roughly reproduces
the same areas of instability seen in the numerical integrations.  We finally
apply this map to areas of semimajor axis phase space that remained unexplored
by the numerical integrations. 

The six degrees of freedom used to initially define planets for these maps
is the set $\{a,e,i,\lambda,\omega, \Omega\}$.  This is the same set of variables
we used for numerical integrations, except here we use the mean longitude ($\lambda$)
instead of the mean anomaly ($M$).  Given initial values of these six variables, we
can define the moments $I_0 = e_{0}^2/2$, $J_0 = i_{0}^2/2$,
$K_0 = 3a_{0}^2/8$, $h_0 = \sqrt{2I_{0}} \cos \omega_0$, and
$k_0 = \sqrt{2I_{0}} \sin \omega_0$, then iterate with our second-order nonplanar
eccentric map, which we obtain after extensive manipulation
of Eqs. (\ref{Iexact})-(\ref{lampr}).  The final map reads:

\beq
h'  =  h_n + \frac{2W_2^{1,0}\sin \lambda_n}{a_3^{2}}
, 
\label{begset}
\eeq

\beq
k'  =  k_n - \frac{2W_2^{1,0}\cos \lambda_n}{a_3^{2}}
,
\eeq

\beq
I'  =  \frac{h'{}^2 + k'{}^2}{2}
,
\eeq

\beq
\omega'  =  \arctan{\frac{k'}{h'}}
,
\eeq

\smallskip

\beq
J_{n+1}  =  \frac{J_n}{1 + \frac{16}{3}a_{3}^{-3}y_{2}\cos[2(\lambda_n-\omega')]\sin[2(\Omega_n-\omega')]}
,
\label{Jint}
\eeq

\smallskip

\beq
I_{n+1}  =  \frac{I' - \frac{16}{3}J_{n+1}y_2 a_3^{-3}\sin\left(2\Omega_n -4\phi' + 2\lambda_n\right)
}{1 + 8\sin\left[2(\lambda_n - \omega')\right]a_3^{-3}W_2^{2,0}} 
,
\label{wackysin}
\eeq

\smallskip

\beq
K_{n+1}  =  I_{n+1} + J_{n+1} - I_n - J_n + K_n 
,
\eeq

\beq
\lambda_{n+1} = \lambda_{n} + \frac{4\pi}{3\epsilon}{\left(\frac{8K_{n+1}}{3}\right)}^{-\frac{1}{2}}
,
\label{M2lambda}
\eeq

\beq
\omega_{n+1}  =  \omega' - \left[4W_2^{2,0}a_3^{-3}\cos\left[2(\lambda_n - \omega')\right] - 2a_3^{-3}\right], 
\eeq

\beq
\Omega_{n+1}  =  \Omega_n - \left[\frac{8}{3}y_2\cos[2(\Omega_n-\omega')]a_3^{-3}\cos
\left[2(\lambda_n - \omega')\right] + 2a_3^{-3}\right] 
.
\label{endset}
\eeq

Some features of Eqs. (\ref{begset})-(\ref{endset}) deserve discussion.
Here, $y_2$ represents $K_1(4/3)$, where $K_1$ is the modified first order Bessel function
of the second kind.  The sine term in Eq. (\ref{wackysin}) is a compact way of
expressing the four double angle terms that result from the application of Eq. (\ref{Iexact}).
Recall that the $W_{n}^{p,q}$ in the mapping equations are constants resulting from the
Taylor expansion of $W$ about $e_{\ast} = 0$ and $i_{\ast} = 0$.  Note that Eqs. (\ref{begset})-(\ref{endset})
contain only the coefficients $W_{2}^{1,0}$ and $W_{2}^{2,0}$, which represent coefficients
of zero inclination terms (see Eq. \ref{Wexpansion}).  The coefficients of the inclination
terms in the expansion do not vanish, but rather were combined into trigonometric arguments
of angle differences, which are manifested in Eqs. (\ref{Jint}), (\ref{wackysin}), and
(\ref{endset}).

Applying the map given by Eqs. (\ref{begset})-(\ref{endset}) to two equal mass planets
provides a null result, unless the Hill parameter is modified suitably.  Because
the equal mass problem deals with two bodies whose orbital parameters change at comparable
and non-negligible rates, the parameters in the map must be scaled accordingly.  Doubling
the Hill parameter satisfies this scaling, so that $\epsilon = 2{\left[({m_1+m_2})/{m_0}\right]}^{1/3}$.
Applying this new Hill parameter to the second-order nonplanar map yields Fig. \ref{fig13}, which
can directly be compared with the instability seen in the numerical integrations for all runs with
$I = 0^{\circ}-15^{\circ}$ with the mapping results. \marginpar{FIG. 13}  The map mirrors the numerical
integrations to a rough extent, providing an estimate for the limit of global chaos
and for instability due to resonance.  The map was run with initial, random mean longitudes,
and was run for just $\approx 1000$ conjunctions.  Also, five times as many initial semimajor axis
values were sampled, but otherwise all initial orbital parameters were the same as the numerical
integrations.  Unstable systems were flagged as those systems whose
relative eccentricity exceeded $1$.  Such systems presented themselves in much less
time (both real time and CPU time) in NLTP's encounter map than in the actual numerical
integrations.  Although NLTP's map provides a fast, convenient way to discover instability in a system,
we find that the results are not accurate enough with respect to our integrations to warrant an
analysis of individual runs.  Hence, we use the map only to explore the gross properties of systems.

In Fig. \ref{fig13}, the map output reproduces the instability arising from
the $5$:$3$ resonance. However, the instability is offset from the actual location of the
resonance.  We assume this offset results from the Taylor expansion of the map around
$i = 0$; as the planet's initial relative inclination increases, the map
is less able to reproduce the actual behavior of the systems.  Applying
the map to relative initial inclinations of $25^{\circ}$ and higher results
illustrates that the map fails to reproduce the instability at the resonances
seen in Fig. \ref{fig5}.  However, for inclinations $\lesssim 15^{\circ}$,
the map can provide a qualitative estimate of the incidence of stability.
Given this conclusion, we now proceed to consider regions of phase space
unexplored by the numerical simulations.

Having established that NLTP's map can produce qualitative features of the
instability seen in the integrations, we ran the map for a wider range of phase
space than that covered by the integrations.  Figure \ref{fig14} shows the
results for a sampling of $6,000$ equally spaced outer planet locations over
the interval [1, 2] for $6$ different inclinations. \marginpar{FIG. 14} One observes
that most systems outward of the limit of global chaos, and the vast majority
of systems outward of the Hill stability limit, display no significant dynamical evolution,
with the notable exceptions of the $5$:$3$ resonance systems and a hint of instability at the $2$:$1$
resonance location.  An equivalent exploration of phase space for the interval [2, 3]
was performed.  Not a single one of those $36,000$ runs became unstable. A final exploration
of [1, 2] phase space was performed for inclinations higher than $15^{\circ}$,
in order to test the robustness of the Taylor expansion about $i = 0$.
The results are presented in Fig. \ref{fig15}. \marginpar{FIG. 15} Inclinations greater than $25^{\circ}$
illustrate nothing discernible, as they fail to reproduce the stable gap located at $1.35$ 
produced by lower inclination trials.

Therefore, encounter maps 1) successfully reproduce the gross properties
of the instabilities seen in the numerical integrations, suggesting
that such maps can be used to model two massive planets, and
2) illustrate that no instability was seen when
$a_2 > 2$, lending support to the concept of Hill stability.

\section{Discussion}

For application to known extrasolar planetary systems, the first question 
is whether multiple giant planets are typically formed in essentially 
coplanar configurations, or instead have significant mutual inclinations. 
Observationally, this is not known, although in some known multiple 
systems stability considerations constrain the maximum mutual inclinations 
of the planets. For example, the two planets in 47 Ursae Majoris cannot 
lie on orbits inclined by more than about $40^\circ$ to each other 
(Laughlin {\it et al.} 2002), while in the more complex 
three-planet system around upsilon Andromeda high inclinations of the 
outer planets appear to be likewise disfavored (Lissauer {\it et al.} 2001;
Rivera 2001). 
These limits, however, while ruling out extreme scenarios such as 
counter rotating orbits, do not preclude mutual inclinations large 
enough to significantly alter the subsequent orbital dynamics. Moreover, 
direct observations of dust debris disks, especially that around $\beta$~Pictoris, 
reveal warps of the order of $10^\circ$ which extend inward to radii 
where giant planets orbit in the Solar System (Wahhaj {\it et al.} 2003).
Although in these cases the observed warps may be the {\em consequence} 
of planet formation rather than a relic of the initial conditions, there 
are also indications that warps might be present---at least at large 
disk radii---at earlier epochs when the protoplanetary disk was still 
gas-rich (Terquem and Bertout 1993; Launhardt and Sargent 2001)

If gravitational scattering is important in the early evolution of 
extrasolar planetary systems (either in initially coplanar configurations, 
or in the inclined systems considered here), then it will influence 
both the orbital radii and eccentricity of the observed post-scattering 
planets. {\em Inward} migration as a consequence of two planet scattering 
is unlikely to explain the existence of hot Jupiters at orbital radii 
$a \leq 0.1~{\rm AU}$, since the progenitor multiple systems would 
themselves need to have formed at small radii where massive planet 
formation is probably inefficient (Bodenheimer {\it et al.} 2000). 
A more promising application is {\em outward} migration from radii 
comparable to that of Jupiter to several tens of AU, where standard 
core accretion models (Pollack {\it et al.} 1996; though see also 
Goldreich {\it et al.} 2004) predict formation timescales that are comparable 
to or longer than typical protoplanetary disk lifetimes. There is 
plausible circumstantial evidence that a population of such planets 
exist, from observations of asymmetries in several debris disks 
(Wilner {\it et al.} 2002; Quillen and Thorndike 2002; Holland {\it et al.} 2003). 
If planet-planet scattering is common in the early evolution of 
planetary systems---as would be required if this is the origin 
of the typically substantial eccentricities of extrasolar planets---then 
the results presented here suggest that the large radii planets could 
be a byproduct of the same process. Large eccentricities of the 
outward scattered planets would then be common, a prediction which 
appears to be consistent with the limited inferences possible from 
existing data.

In addition to the (small) fraction of known extrasolar planets that 
have been found at surprisingly small orbital radii, a large fraction 
of the planets further out have significantly eccentric orbits. For 
massive planets outside the hot Jupiter zone, eccentricity is 
distributed approximately uniformly in the range $0 < e < 0.7$ 
(Marcy {\it et al.} 2003), with a smaller number of outliers at even 
higher eccentricity. This is certainly qualitatively consistent with a 
high frequency of planet-planet scattering, though whether multiple 
planet formation is {\em required} to yield significantly non-circular 
orbits depends on the rather uncertain issue of whether a single 
planet's eccentricity can be excited as a consequence of interaction 
with the protoplanetary disk (Artymowicz 1992; Papaloizou {\it et al.} 
2001; Goldreich and Sari 2003; Ogilvie and Lubow 2003). Encouragingly 
for the multiple planet scenarios (Rasio and Ford 1996; Ford {\it et al.} 2001; 
Marzari and Weidenschilling 2002), two-planet integrations that include a 
realistic range of planet masses (Ford {\it et al.} 2003) provide a 
better quantitative match to the observed eccentricity distribution than 
did prior simulations assuming equal mass planets. 

Overall then, the available evidence is consistent with a high frequency 
of gravitational scattering in initially multiple massive planet systems. 
The calculations presented here, however, confirm that gravity is 
unlikely to be the only significant agent at work. If planets are 
formed with extremely small orbital separations, then interesting 
dynamical behavior (excitation of eccentricity and/or ejection of 
one of the planets) is inevitable, but the timescale is typically 
short. A significant fraction of these eventually unstable systems develop 
instability in $10^2-10^5$ years, a shorter timescale than the typical disk
dissipation timescale (Hartigan {\it et al.} 1990; Wolk and Walter 1996). 
This implies that purely gravitational scattering experiments are 
physically inconsistent---in reality either such unstable initial conditions 
could not arise, or additional forces such as torques from the 
remnant protoplanetary disk would have to play a role. If, alternatively, 
multiple planets are formed with larger separations that exceed the 
global chaos limit, then as we have seen the most interesting 
dynamical behavior is restricted to the vicinity of low-order mean motion 
resonances. If planets form at random locations with the protoplanetary 
disk, then only a few percent of all systems would satisfy this condition. 
In fact, approximately ten percent of known extrasolar planets are found 
within multiple systems, and these planets have a similar distribution 
of eccentricities to those planets with no (known) companions (Marcy {\it et al.} 
2003). Several of the multiple systems are either definitely or probably 
in resonance. Taken as a whole, we believe that these properties are 
consistent with a scenario in which {\em both} disk-driven migration 
and planet-planet interactions play critical roles in explaining the 
origin of extrasolar planet eccentricities (e.g. Snellgrove {\it et al.} 
2001; Lee and Peale 2003; Chiang 2003), although quantifying the 
statistical properties of the resulting planetary systems within such 
models remains difficult.

\section{Conclusions}

We have explored the dynamics of two close massive planets on inclined, nearly circular
orbits using both analytical and numerical techniques.  Our findings are summarized as follows:
First, our main result is that significant outward radial migration of EGPs through 
gravitational scattering alone is possible and occurs either inside the region of global chaos
or at the first, second, or third mean motion resonances (especially the $2$:$1$ and
$5$:$3$ resonances).  The few cases of
migration that do not occur at such resonances migrate at the next lowest-order
resonance locations.  Scattering can also lead to some inward migration, but for two
equal masses, the degree of inward migration is constrained by energetic arguments to be
modest (Lin and Ida 1999; Papaloizou and Terquem 2001; Adams and Laughlin 2003).

Our ancillary results are 1) the Hill stability limit for arbitrarily inclined circular
orbits is independent of mass to first order, and is given by Eq. (\ref{finali}).
This result restricts the migratory behavior of EGPs, placing constraints on the
initial semimajor axis range where a system can be dynamically excited.
2) For almost circular, arbitrary inclined orbits, we obtained analytical
formulas for the secular evolution of some orbital parameters expressed in
terms of the masses and initial semimajor axes only (Eqs. \ref{i2t}-\ref{i1t},
\ref{node1var}, and \ref{node2var}). 3) Laplace-Lagrange
secular theory fails to trace the eccentricity evolution two planets, both
on initially circular orbits.  4) Libration widths computed with
nonzero values of mass, eccentricity, and inclination for {\it each} planet (Eq. \ref{lib8}) 
appear to be viable tracers of the dynamical reach of resonances.  5) Encounter maps may
be used to study the gross properties of the dynamics of EGP systems with a
suitable scaling of the Hill parameter.

\bigskip
\bigskip

\newpage

\section*{Acknowledgments}

We wish to thank two anonymous referees for their insights and helpful suggestions,
Larry Esposito, Glen Stewart and the rest of the Colorado Rings Group for vital guidance, 
Kevin Rauch and Doug Hamilton for use of their HNbody code, Henry Tufo for extensive use of 
the Hemisphere 64-cluster of computers at the Colorado BP Center for Visualization, Eugene Chiang 
for a productive discussion, and Juri Toomre for the use of the Laboratory for Computational Dynamics.

\bigskip

This paper is based upon work supported by NASA under Grant NAG5-13207 issued through the Office of 
Space Science.  Computer time at the Colorado BP Center for Visualization was provided by equipment 
purchased under NSF ARI Grant \#CDA-9601817 and NSF sponsorship of the National Center for 
Atmospheric Research.

\newpage

\section*{References}

\noindent{}Adams, F.C., Laughlin, G., 2003. Migration and dynamical relaxation in crowded systems of giant planets. Icarus 163, 290-306.

\bigskip

\noindent{}Agnor, C.B., Ward, W.R., 2002.  Damping of Terrestrial Planet Eccentricities by Density Wave Interactions with a Remnant Gas Disk.
DDA Meeting \#33, \#07.12, 938.

\bigskip

\noindent{}Artymowicz, P., 1992.  Dynamics of binary and planetary-system interaction with disks - Eccentricity changes.  Astronom. Soc. Pac.
104, 769-774.

\bigskip

\noindent{}Beagu\'{e}, C., Michtchenko, T.A., 2003. Modelling the high-eccentricity planetary three-body problem. Application to the GJ876 planetary system. Mon. Not. R. Astron. Soc.  341, 760-770.

\bigskip

\noindent{}Bodenheimer, P., Hubickyj, O., Lissauer, J.J., 2000.  Models of the in Situ Formation of Detected Extrasolar Giant Planets.  Icarus 143, 2-14.

\bigskip

\noindent{}Bouchy, F., Pont, F., Santos, N.C., Melo, C., Mayor, M., Queloz, D., Udry, S., 2004.  Two new "very hot Jupiters" among the OGLE transiting candidates.  preprint (astro-ph/0404264).

\bigskip

\noindent{}Brouwer, D. and Clemence, G.M., 1961. Methods of Celestial Mechanics.  Academic Press, New York, London.

\bigskip

\noindent{}Carruba, V., Burns, J.A., Nicholson, P.D., 2002. On the Inclination Distribution of the Jovian Irregular Satellites. Icarus 158, 434-449.

\bigskip

\noindent{}Champenois, S., Vienne, A., 1999a.  The Role of Secondary Resonances in the Evolution of the Mimas-Tethys System. Icarus 140, 106-121.

\bigskip

\noindent{}Champenois, S., Vienne, A., 1999b.  Chaos and secondary resonances in the mimas-tethys system. Celest. Mech. 74, 111-146.

\bigskip

\noindent{}Charbonneau, D., Brown, T.M., Latham, D.W., Mayor M., 2000.  Detection of Planetary Transits Across a Sun-like Star. Astrophys. J.
529, L45-L48.

\bigskip

\noindent{}Chiang, E.I., 2003.  Excitation of Orbital Eccentricities by Repeated Resonance Crossings: Requirements.  Astrophys. J. 584, 465-471.

\bigskip

\noindent{}Chiang, E.I., Fischer, D., Thommes, E., 2002.  Excitation of Orbital Eccentricities of Extrasolar Planets by Repeated
Resonance Crossings. Astrophys. J. 564, L105-L109.

\bigskip

\noindent{}Duncan, M., Quinn, T., Tremaine, S., 1989. The long-term evolution of orbits in the solar system - A mapping approach. Icarus 82, 402-418.

\bigskip

\noindent{}Ford, E.B., Havlickova, M., Rasio, F.A., 2001. Dynamical Instabilities in Extrasolar Planetary Systems Containing Two Giant Planets.  Icarus 150, 303-313.

\bigskip

\noindent{}Ford, E.B., Rasio, F.A., Yu, K., 2003.  Dynamical Instabilities in Extrasolar Planetary Systems.  Preprint (astro-ph/0210275).

\bigskip

\noindent{}Gladman, B., 1993.  Dynamics of systems of two close planets. Icarus 106, 247-263.

\bigskip

\noindent{}Goldreich, P., Lithwick, Y., Sari, R., 2004.  Formation of Kuiper Belt Binaries.  Preprint (astro-ph/0208490)

\bigskip

\noindent{}Goldreich, P., Sari, R., 2003. Eccentricity Evolution for Planets in Gaseous Disks.  Astrophys. J. 585, 1024-1037.

\bigskip

\noindent{}Goldreich, P., Tremaine, S., 1980.  Disk-satellite interactions.  Astrophys. J. 241, 425-441.

\bigskip

\noindent{}Hartigan, P., Hartmann, L., Kenyon, S.J., Strom, S.E., Skrutskie, M.F., 1990.   Correlations of optical and infrared
excesses in T Tauri stars.  Astrophys. J. 354, L25-L28.

\bigskip

\noindent{}Hen\'{o}n, M., Petit J., 1986.  Series expansion for encounter-type solutions of Hill's problem. Celest. Mech.  38, 67-100.

\bigskip

\noindent{}Henrard, J., Watanabe, N., Moons, M. 1995. A Bridge between Secondary and Secular Resonances inside the Hecuba Gap.  Icarus 115, 336-346.

\bigskip

\noindent{}Henry, G.W., Marcy, G.W., Butler, R.P., Vogt, S.S., 2000.  A Transiting ``51 Peg-like'' Planet.  Astrophys. J. 529, L41-L44.

\bigskip

\noindent{}Holland, W.S., Greaves, J.S., Dent, W.R.F., Wyatt, M.C., Zuckerman, B., Webb, R.A., McCarthy, C., Coulson, I.M., Robson, E.I., Gear, W.K., 2003.  Submillimeter Observations of an Asymmetric Dust Disk around Fomalhaut. Astrophys. J. 582, 1141-1146.

\bigskip

\noindent{}Hill, G.W., 1878.  Researches in the Lunar Theory. American J. Math. 1, 5-26, 129-147, 245-260.

\bigskip

\noindent{}Holman, M., Touma, J., Tremaine, S., 1997.  Chaotic variations in the eccentricity of the planet
orbiting 16 Cygni B.  Nature 386, 254-256.

\bigskip

\noindent{}Julian, W.H., Toomre, A., 1966. Non-Axisymmetric Responses of Differentially Rotating Disks of Stars.  Astrophys. J. 146, 810-830.

\bigskip

\noindent{}Kato, S., 1983. Low-frequency, one-armed oscillations of Keplerian gaseous disks.  Astron. Soc. Jap. 35, 249-261.

\bigskip

\noindent{}Kaula, W.M., 1962.  Development of the lunar and solar disturbing functions for a close satellite.  Astron. J. 67, 300-303.

\bigskip

\noindent{}Konacki, M., Torres, G., Jha, S., Sasselov, D., 2003.  An extrasolar planet that transits the disk of its parent star. Nature 421, 507-509.

\bigskip

\noindent{}Konacki, M., Torres, G., Sasselov, D.D., Pietrzynski, G.,  Udalski, A., Jha, S., Ruiz, M.T., Gieren, W., Minniti, D., 2004.  A transiting extrasolar giant planet around the star OGLE-TR-113.  Preprint (astro-ph/0404541).

\bigskip

\noindent{}Laughlin, G., Chambers, J., Fischer, D., 2002.  A Dynamical Analysis of the 47 Ursae Majoris Planetary System.  Astrophys. J. 579, 455-467.

\bigskip

\noindent{}Launhardt, R., Sargent, A.I., 2001.  A Young Protoplanetary Disk in the Bok Globule CB 26?  Astrophys. J. 562, L173-L175.

\bigskip

\noindent{}Lee, M.H., Peale, S. J., 2002.  Dynamics and Origin of the 2:1 Orbital Resonances of the GJ 876 Planets.  Astrophys. J.  567, 596-609.

\bigskip

\noindent{}Lin, D.N.C., Bodenheimer, P., Richardson, D.C., 1996.  Orbital migration of the planetary companion of 51 Pegasi to its present location.  Nature 380, 606-607.

\bigskip

\noindent{}Lin, D.N.C., Ida, S., 1997.   On the Origin of Massive Eccentric Planets.  Astrophys. J. 477, 781-791.

\bigskip

\noindent{}Lin, D.N.C., Papaloizou, J.C.B., 1986.  On the tidal interaction between protoplanets and the protoplanetary disk. III - Orbital migration of protoplanets.  Astrophys. J. 309, 846-857.

\bigskip

\noindent{}Lissauer, J.J., Rivera, E.J., 2001.  Stability Analysis of the Planetary System Orbiting $\nu$ 
Andromedae. II. Simulations Using New Lick Observatory Fits.  Astrophys. J. 554, 1141-1150.

\bigskip

\noindent{}Lubow, S.H., Ogilvie, G.I., 2001.  Secular Interactions between Inclined Planets and a Gaseous Disk.  Astrophys. J. 560, 997-1009.

\bigskip

\noindent{}Lyubarskij, Y.E., Postnov, K.A., Prokhorov, M.E., 1994.  Eccentric Accretion Discs. Mon. Not. R. Astron. Soc.  266, 583-596.

\bigskip

\noindent{}Marchal, C., Bozis, G., 1982. Hill Stability and Distance Curves for the General Three-Body Problem. Celest. Mech. 26:  311-333.

\bigskip

\noindent{}Marcy, G.W., Butler, R.P., Fischer, D.A., Vogt, S.S., 2003.  Properties of Extrasolar Planets. Scientific Frontiers in Research on Extrasolar Planets 1-16.

\bigskip

\noindent{}Marzari, F., Weidenschilling, S.J., 2002.  Eccentric Extrasolar Planets: The Jumping Jupiter Model.  Icarus 156, 570-579.

\bigskip

\noindent{}Murray, C.D., and Dermott, S.F., 1999. Solar System Dynamics. Cambridge Univ. Press, Cambridge, UK.

\bigskip

\noindent{}Nagasawa, M., Lin, D.N.C., Ida, S., 2003. Eccentricity Evolution of Extrasolar Multiple Planetary Systems Due to the Depletion of Nascent Protostellar Disks.  Astrophys. J. 586, 1374-1393.

\bigskip

\noindent{}Namouni, F., Luciani, J.F., Tabachnik, S., Pellat, R., 1996. A mapping approach to Hill's distant encounters: application to the stability of planetary embryos. Astron. Astrophys. 313, 979-992.

\bigskip

\noindent{}Ogilvie, G.I., 2001. Non-linear fluid dynamics of eccentric discs.  Mon. Not. R. Astron. Soc. 325, 241-248.

\bigskip

\noindent{}Ogilvie, G.I., Lubow, S.H., 2003.  Saturation of the Corotation Resonance in a Gaseous Disk.  Astrophys. J. 587, 398-406.

\bigskip

\noindent{}Papaloizou, J.C.B., Nelson, R.P., Masset, F., 2001.  Orbital eccentricity growth through disc-companion tidal interaction. Astron. Astrophys.  366, 263-275.

\bigskip

\noindent{}Papaloizou, J.C.B., Terquem, C., 2001.  Dynamical relaxation and massive extrasolar planets. Mon. Not. R. Astron. Soc.  325, 221-230.

\bigskip

\noindent{}Peale, S.J., 1986. Orbital resonances, unusual configurations and exotic rotation states among planetary satellites.  In: Burns, J.A., Matthews, M.S. (Eds.), Satellites, University of Arizona Press, Tucson, pp. 159-223. 

\bigskip

\noindent{}Peale, S.J., 1999. Origin and Evolution of the Natural Satellites. Annu. Rev. Astron. Astrophys. 37: 533-602.

\bigskip

\noindent{}Pollack, J. B., Hubickyj, O., Bodenheimer, P., Lissauer, J.J., Podolak, M., Greenzweig, Y., 1996.  Formation of the Giant Planets by Concurrent Accretion of Solids and Gas.  Icarus 124, 62-85.

\bigskip

\noindent{}Quillen, A. C., Thorndike, S., 2002.  Structure in the {$\epsilon$} Eridani Dusty Disk Caused by Mean Motion Resonances with a 0.3 Eccentricity Planet at Periastron.  Astrophys. J. 578, L149-L152.

\bigskip

\noindent{}Rasio, F.A., Ford, E.B., 1996.  Dynamical instabilities and the formation of extrasolar planetary systems.  Science 274, 954-956.

\bigskip

\noindent{}Rice, W.K.M., Armitage, P.J., Bonnell, I.A., Bate, M.R., Jeffers, S.V., Vine, S.G., 2003.  Substellar companions and isolated
planetary-mass objects from protostellar disc fragmentation. Mon. Not. R. Astron. Soc. 346, L36-L40.

\bigskip

\noindent{}Roig, F., Simula, A., Ferraz-Mello, S., Tsuchida, M., 1998.  The high-eccentricity asymmetric expansion
of the disturbing function for non-planar resonant problems.  Astron. Astrophys. 329, 339-349.

\bigskip

\noindent{}Schneider, J., 2004, http://cfa-www.harvard.end/planets/catalog.html

\bigskip

\noindent{}Sinclair, A.T., 1972. On the origin of the commensurabilities amongst the satellites of Saturn. Mon. Not. R. Astron. Soc. 160, 169-187.

\bigskip

\noindent{}Snellgrove, M. D., Papaloizou, J. C. B., Nelson, R. P., 2001.  On disc driven inward migration of resonantly coupled planets with application to the system around GJ876.  Astron. Astrophys. 374, 1092-1099. 

\bigskip

\noindent{}Terquem, C., Bertout, C., 1993.  Tidally-Induced WARPS in T-Tauri Disks - Part One - First Order Perturbation Theory.  Astron. Astrophys. 274, 291-303.

\bigskip

\noindent{}Terquem, C., Bertout, C., 1996.  Tidally induced WARPS in T Tauri discs - II. A parametric study of spectral energy distributions.  Mon. Not. R. Astron. Soc. 279, 415-428.

\bigskip

\noindent{}Thommes, E.W., Lissauer, J.J., 2003.  Resonant Inclination Excitation of Migrating Giant Planets.  Astrophys. J. 597, 566-580.

\bigskip

\noindent{}Veras, D., Armitage, P.J., 2004.  Outward migration of extrasolar planets to large orbital radii.  Mon. Not. R. Astron. Soc. 347, 613-624.

\bigskip

\noindent{}Wahhaj, Z., Koerner, D. W., Ressler, M. E., Werner, M. W., Backman, D. E., Sargent, A. I., 2003.  The Inner Rings of $\beta$ Pictoris.  Astrophys. J. 584, L27-L31.

\bigskip

\noindent{}Weidenschilling, S.J., Chapman, C.R., Davis, D.R., Greenberg, R., 1984. Ring particles - Collisional interactions and physical nature.  In: 
	   Greenberg, R, Brahic, A. (Eds.), Planetary Rings, University of Arizona Press, Tuscon, pp. 367-415.

\bigskip

\noindent{}Wilner, D. J., Holman, M. J., Kuchner, M. J., Ho, P. T. P., 2002.  Structure in the Dusty Debris around Vega.  Astrophys. J. 569, L115-L119.

\bigskip

\noindent{}Wisdom, J., 1980. The resonance overlap criterion and the onset of stochastic behavior in the restricted three-body problem.
Astron. J. 85, 1122-1133.

\bigskip

\noindent{}Wisdom, J., 1987. Urey Prize Lecture - Chaotic dynamics in the solar system. Icarus 72, 241-275.

\bigskip

\noindent{}Wolk, S.J., Walter, F.M., 1996.  A Search for Protoplanetary Disks Around Naked T Tauri Stars.  Astron. J. 111, 2066-2067.

\bigskip

\noindent{}Yu, Q., Tremaine, S., 2001. Resonant Capture by Inward-migrating Planets. Astronom. J. 121, 1736-1740.

\newpage

\begin{table}[h]
\begin{center}
	\renewcommand{\arraystretch}{1.75}
\begin{tabular}{c || c | c ||  c | c || c | c || c | c || c | c || c | c || c |}  \cline{2-14}
                       & \multicolumn{2}{c ||}{\bf 8:5} & \multicolumn{2}{c ||}{\bf 5:3} & \multicolumn{2}{c ||}{\bf 7:4}   & \multicolumn{2}{c ||}{\bf 2:1} & \multicolumn{2}{c ||}{\bf 5:2} & \multicolumn{2}{c ||}{\bf 3:1} & \% accounted  \\ \cline{2-13}
                       & \multicolumn{2}{c ||}{\bf $a_2=1.368$}  & \multicolumn{2}{c ||}{\bf $a_2=1.406$} & \multicolumn{2}{c ||}{\bf $a_2=1.452$}   & \multicolumn{2}{c ||}{\bf $a_2=1.587$} & \multicolumn{2}{c ||}{\bf $a_2=1.842$} & \multicolumn{2}{c ||}{\bf $a_2=2.080$} & for by \\ \cline{2-13}
                       & $\pm 1 w$ &  $\pm 2 w$  & $\pm 1 w$  & $\pm 2 w$  & $\pm 1 w$  & $\pm 2 w$ & $\pm 1 w$  & $\pm 2 w$ & $\pm 1 w$ & $\pm 2 w$  & $\pm 1 w$  & $\pm 2 w$  & resonances\\ \hline\hline
$I = 0^{\circ}$    &  &  &  &  &  &  &  &  &  &  &  &  &  \\ \hline
$I = 5^{\circ}$    &  &  &  &  &  &  &  &  &  &  &  &  & \\ \hline
$I = 10^{\circ}$   & 1 & 1 & 10 & 11 &  &  &  &  &  &  &  &  & $100\% \ls (12/12)$\\ \hline
$I = 15^{\circ}$   & 0 & 3 & 12 & 13 & 1 & 1 &  &  &  &  &  &  & $100\% \ls (17/17)$\\ \hline
$I = 20^{\circ}$   & 1 & 5 & 6  & 8  & 3 & 3 & 3 & 5 &  &  &  &  & $100\% \ls (19/19)$\\ \hline
$I = 25^{\circ}$   &  &  & 6 & 8 & 2 & 2 & 13 & 15 &  &  &  &  & $100\% \ls (24/24)$\\ \hline
$I = 30^{\circ}$   &  &  &  &  &  &  & 9 & 11 & 0 & 0 &  &  & $92\% \ls (11/12)$\\  \hline
$I = 35^{\circ}$   &  &  &  &  &  &  & 1 & 2 & 0 & 0 & 0 & 0 & $67\% \ls (2/3)$\\ \hline\hline
    {\bf Total \%}   & \multicolumn{2}{c ||}{$10\% \ls (9/87)$} & \multicolumn{2}{c ||}{$46\% \ls (40/87)$} & \multicolumn{2}{c ||}{$7\% \ls (6/87)$} & \multicolumn{2}{c ||}{$38\% \ls (33/87)$} & \multicolumn{2}{c ||}{}  & \multicolumn{2}{c ||}{} & $98\% \ls (85/87)$\\ \hline
\end{tabular}
\caption{\label{vars}  Summary of the number of dynamically rigorous but stable systems affected by every 1st, 2nd, and 3rd order
resonance in the semimajor axis range sampled in the simulations.  The effect of each resonance is measured by its libration
width, and twice that value, denoted respectively by $1w$ and $2w$.  Initial relative inclinations are given by the leftmost
column.}
\end{center}
\label{Ta:resta}
\end{table}

\newpage

\begin{table}[hb]
\begin{center}
	\renewcommand{\arraystretch}{1.75}
\begin{tabular}{ c     |        c        |        c          |        c           }
     Initial Variables & Final Variables & Mapping Variables & Other Variables           \\ \hline\hline
     $m_0, m_1, m_2$   & & & $\epsilon = {\left[\frac{m_1+m_2}{m_0}\right]}^{\frac{1}{3}}$\\ \hline
     $a_1, a_2$        & $a_r = \frac{a_1-a_2}{\epsilon a_0}$ &  $K = \frac{3}{8} a_{r}^2$ &  $a_2 < a_0 < a_1$ \\ \hline
     $e_1, e_2$        & $e_r = \frac{e_1-e_2}{\epsilon}$     &  $I = \frac{e_{r}^2}{2}$ & $e_{\ast} = \frac{e_r}{a_r}$ \\ \hline
     $i_1, i_2$        & $i_r = \frac{i_1-i_2}{\epsilon}$     &  $J = \frac{i_{r}^2}{2}$ & $i_{\ast} = \frac{i_r}{a_r}$ \\ \hline
     $\lambda_1, \lambda_2$ &   $\lambda_r = \lambda_1 - \lambda_2$ &                    & \\ \hline
     $\Omega_1, \Omega_2$  &  $\Omega_r = \Omega_1 - \Omega_2$ &                                   & \\ \hline
     $\omega_1, \omega_2$  &  $\omega_r = \omega_1 - \omega_2$ &  $\substack{h = e_r\cos\omega_r \\ k = e_r\sin\omega_r}$ &  $\theta_r = \omega_r - \lambda_r$\\ \hline
\end{tabular}
\caption{\label{vars}  Summary of variables used in the NLTP map. }
\end{center}
\label{Ta:vars}
\end{table}

\newpage

\section*{Figure Captions}

\begin{figure}

\centerline{\psfig{figure=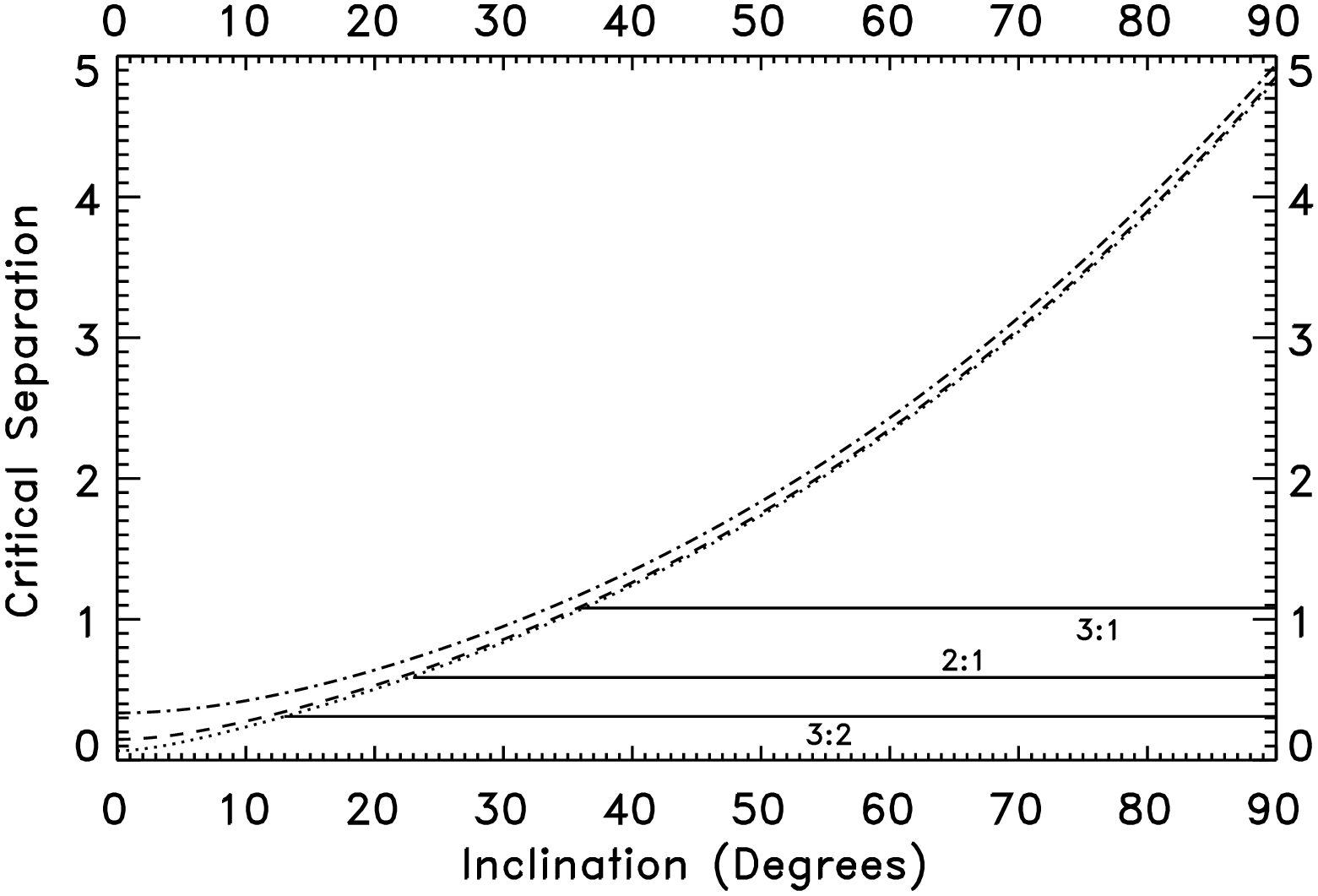,width=7.0truein,height=4.0truein}}
\caption{}
\label{fig1}
\end{figure}

Figure \ref{fig1}:  Critical Hill Separation as a function of relative initial inclination
when $a_1 = 1$.  The dot-dashed, dashed, and dotted lines correspond to
two planets with masses $10^{-3} M_{\odot}$, $10^{-4}  M_{\odot}$,
and $10^{-5}  M_{\odot}$ respectively.  The solid lines indicate the
locations of major first and second order resonances.

\bigskip

\begin{figure}

\centerline{\psfig{figure=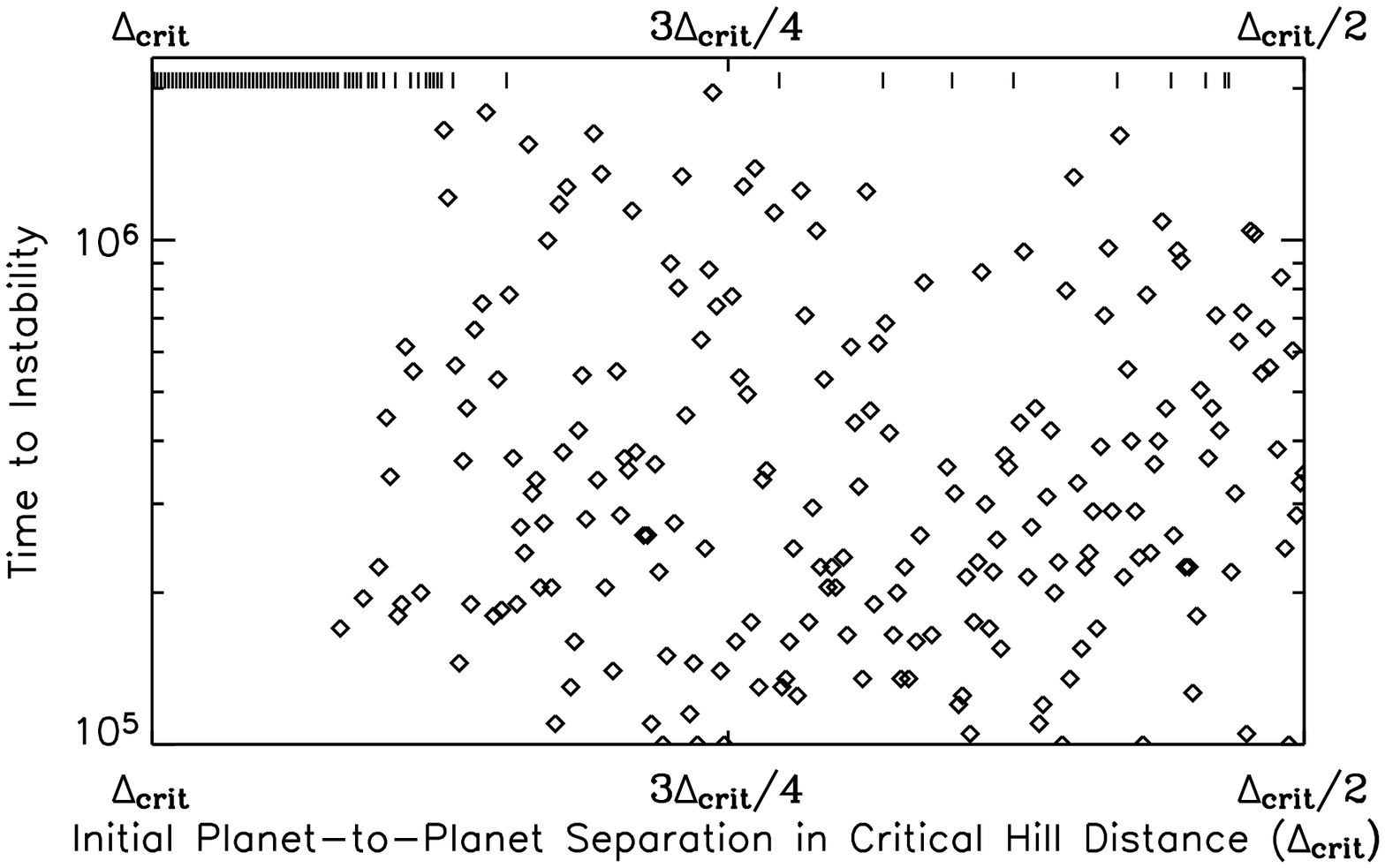,width=7.0truein,height=4.0truein}}
\caption{}
\label{fig2}
\end{figure}

Figure \ref{fig2}:  Time to instability for initially unstable coplanar systems.  Diamonds
mark unstable systems, and bars mark stable systems, all of which were stopped after
$2 \times 10^6$.  These time units correspond to years for an inner planet with a semimajor 
axis of $1$ AU.

\bigskip

\begin{figure}

\centerline{\psfig{figure=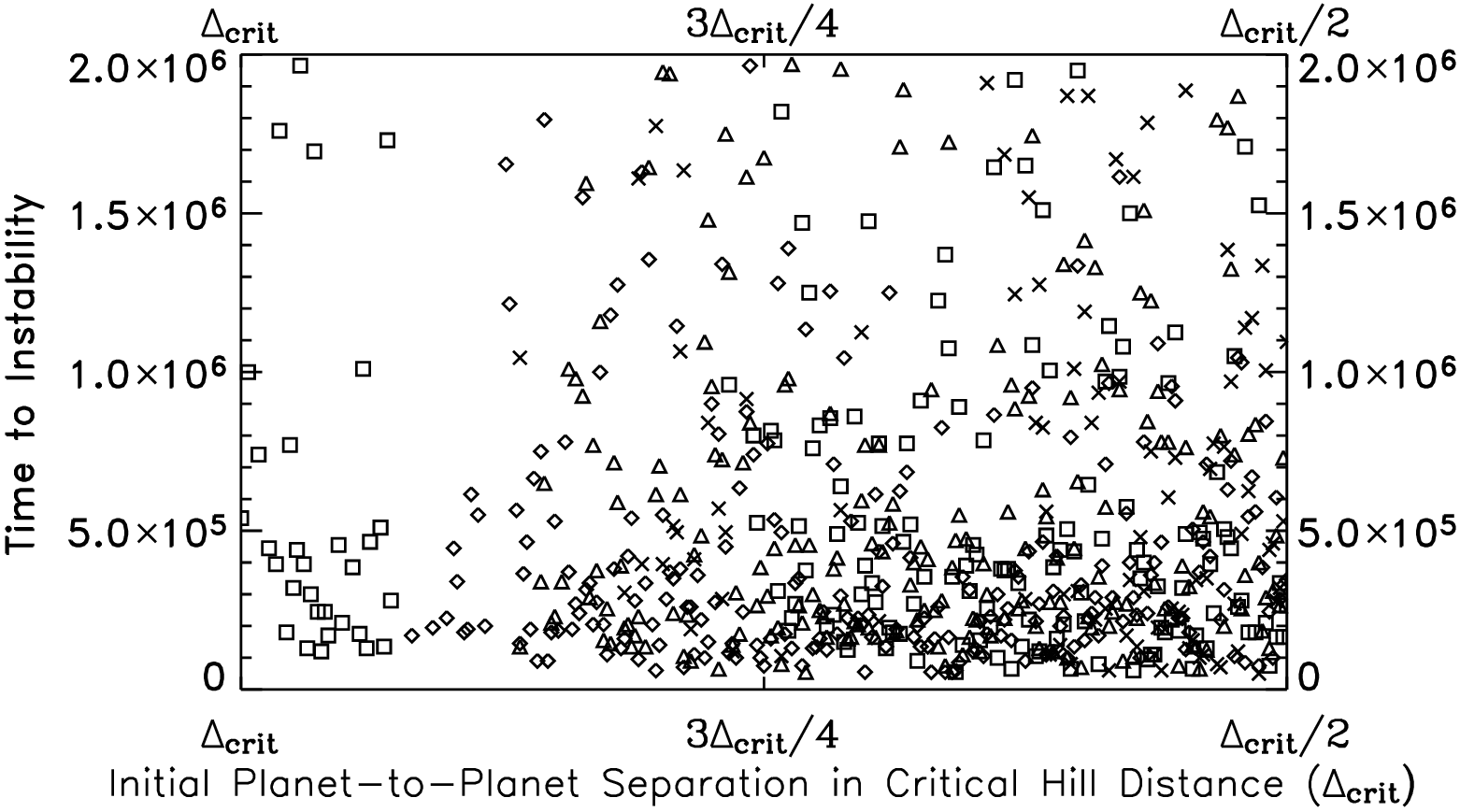,width=7.0truein,height=4.0truein}}
\caption{}
\label{fig3}
\end{figure}

Figure \ref{fig3}:  Same as Fig. \ref{fig2} for systems with initial relative
inclinations of $0^{\circ}$ (diamonds), $5^{\circ}$ (triangles), $10^{\circ}$ (squares),
and $15^{\circ}$ (crosses).  Stable systems are not shown.

\bigskip

\begin{figure}

\centerline{\psfig{figure=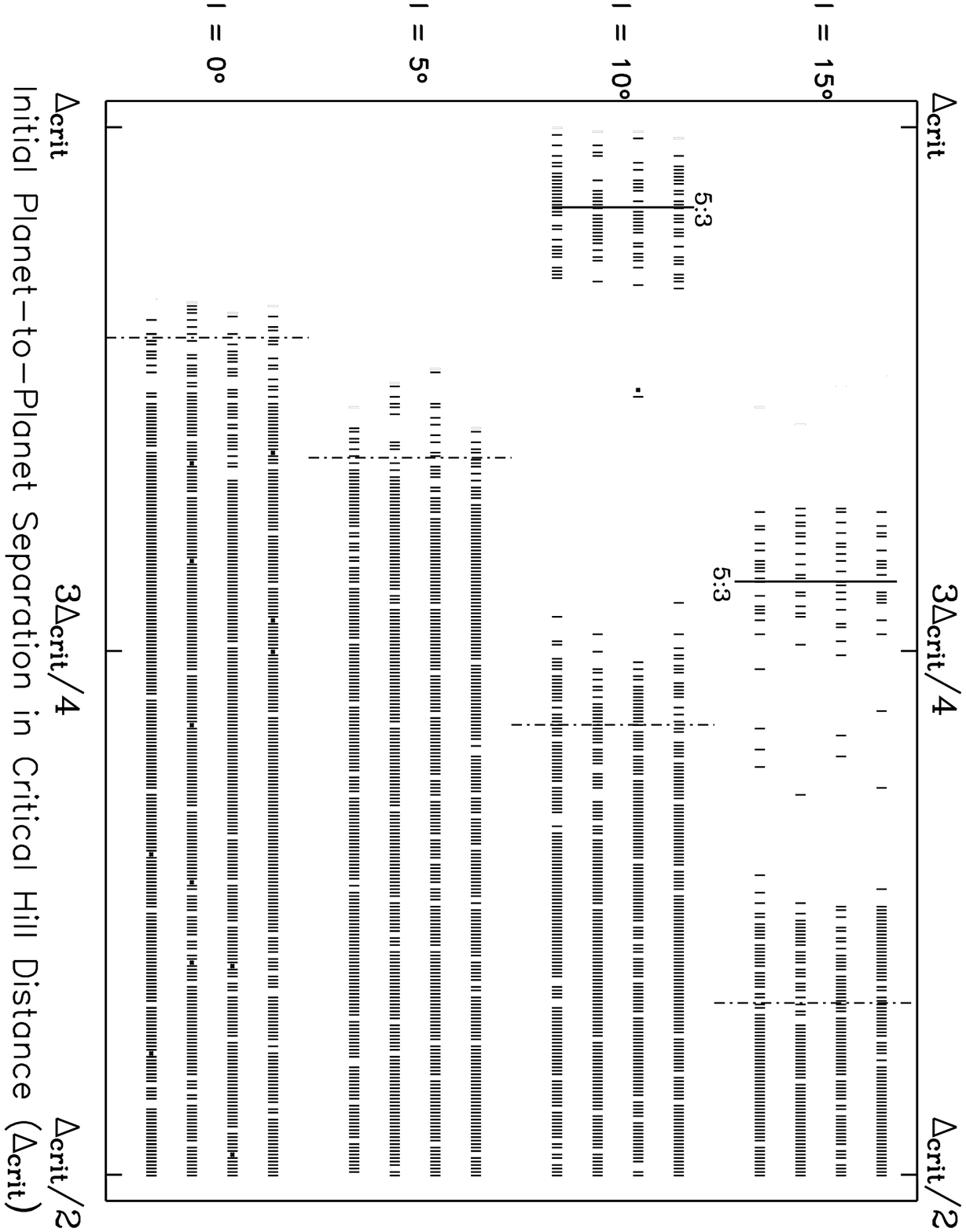,width=7.5truein,height=9.0truein}}
\caption{}
\label{fig4}
\end{figure}

Figure \ref{fig4}:  Record of instability.  Shown are four sets of four rows of bars,
each set corresponding to a different initial relative inclination value, and
each row corresponding to a different trial for given initial values of $a_2$ and $I$.  
Each bar represents an unstable simulation; absence of a bar indicates
a stable simulation; an ``x'' represents simulations that the HNbody integrator
failed to complete because it could not achieve the desired accuracy.  The horizontal
axis provides the initial planet separation, where the critical separation $\Delta_{crit}$ is
computed from Eq. (\ref{finali}).  The dot-dashed vertical lines
represent global chaos limits ($\propto \mu^{2/7}$)
conjectured by Wisdom (1980) for equal mass planets on circular
planar orbits.  The coefficient of $\mu^{2/7}$ was fit for the $I=0^{\circ}$ case, and 
applied to the other cases.  The 5:3 resonance location is displayed with solid 
vertical lines.

\bigskip

\begin{figure}

\centerline{\psfig{figure=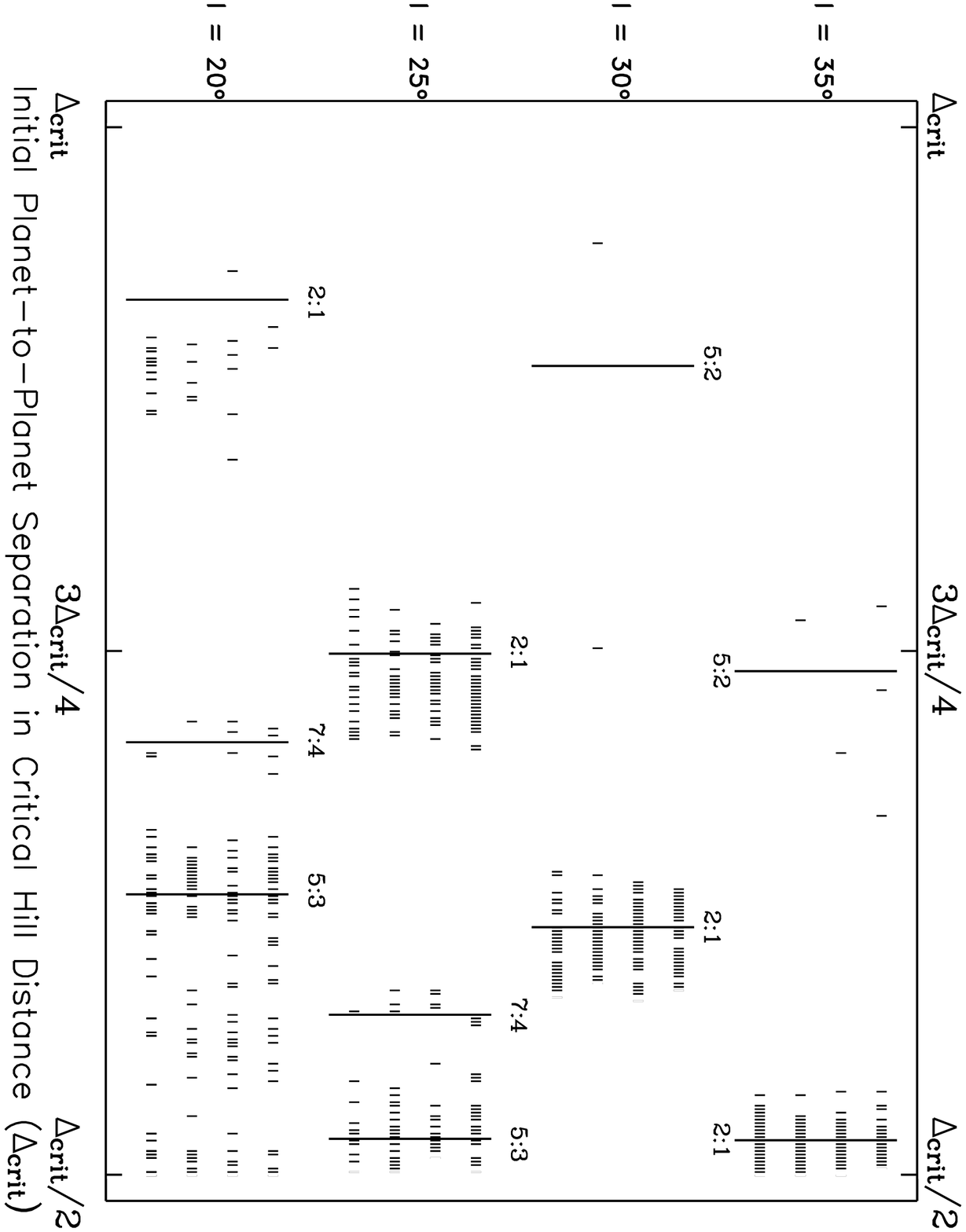,width=7.5truein,height=9.0truein}}
\caption{}
\label{fig5}
\end{figure}

Figure \ref{fig5}: Continuation of Fig. \ref{fig4} for initial relative inclination
values of $20^{\circ}$ to $35^{\circ}$.  All first, second, and third order
nominal resonances locations in the sampled regions are shown.  The region
of global chaos extends beyond (is less than) $\Delta_{crit}/2$ for most of the inclinations
sampled here, and hence is not seen except around $\Delta_{crit}/2$ for $I = 20^{\circ}$.

\bigskip

\begin{figure}

\centerline{\psfig{figure=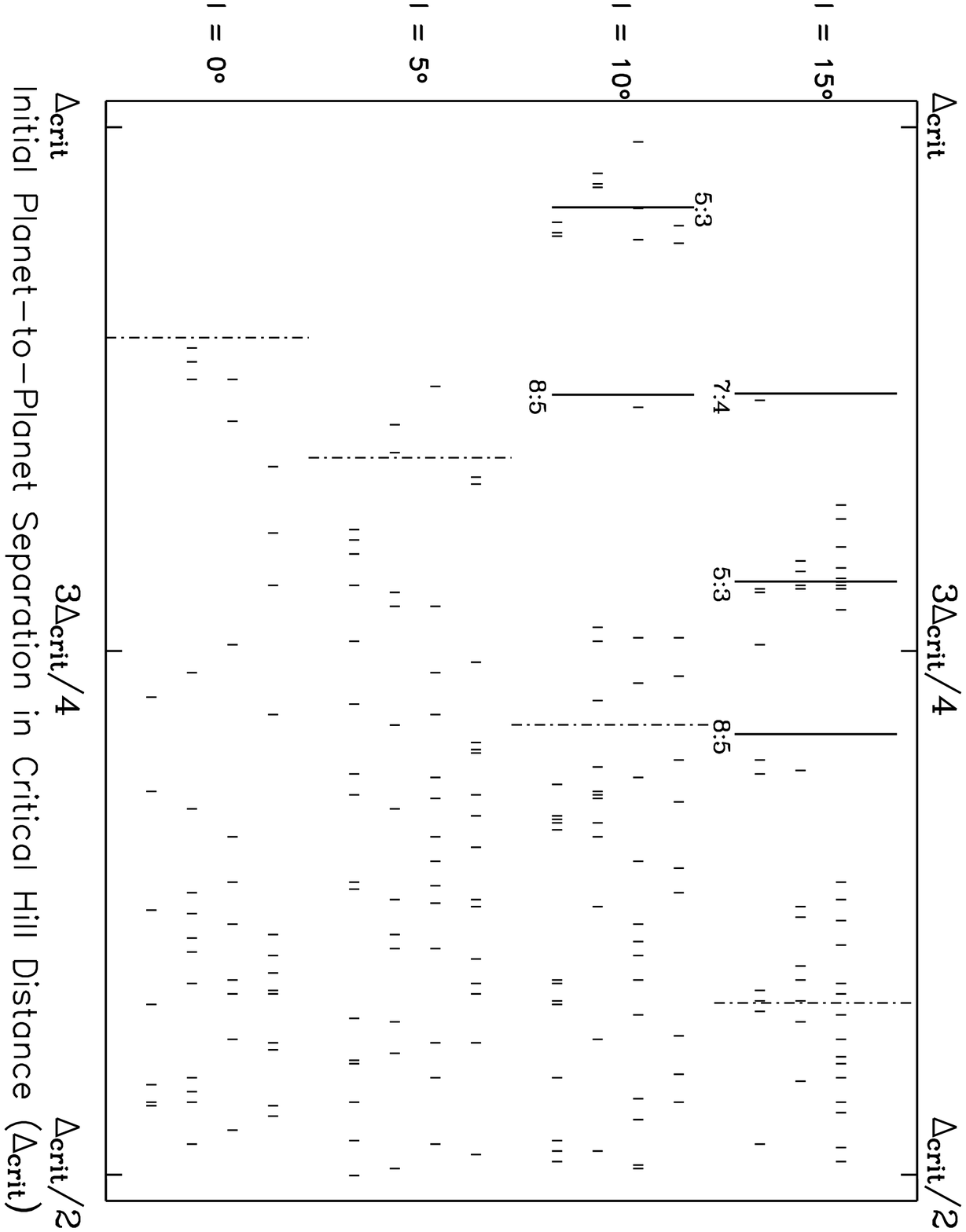,width=7.5truein,height=9.0truein}}
\caption{}
\label{fig6}
\end{figure}

Figure \ref{fig6}: Same as Fig. \ref{fig4} {\it except} that bars represent 
stable systems where significant migration (semimajor axis of either planet
differing by at least 20\% of its initial value) occurs.

\bigskip

\begin{figure}

\centerline{\psfig{figure=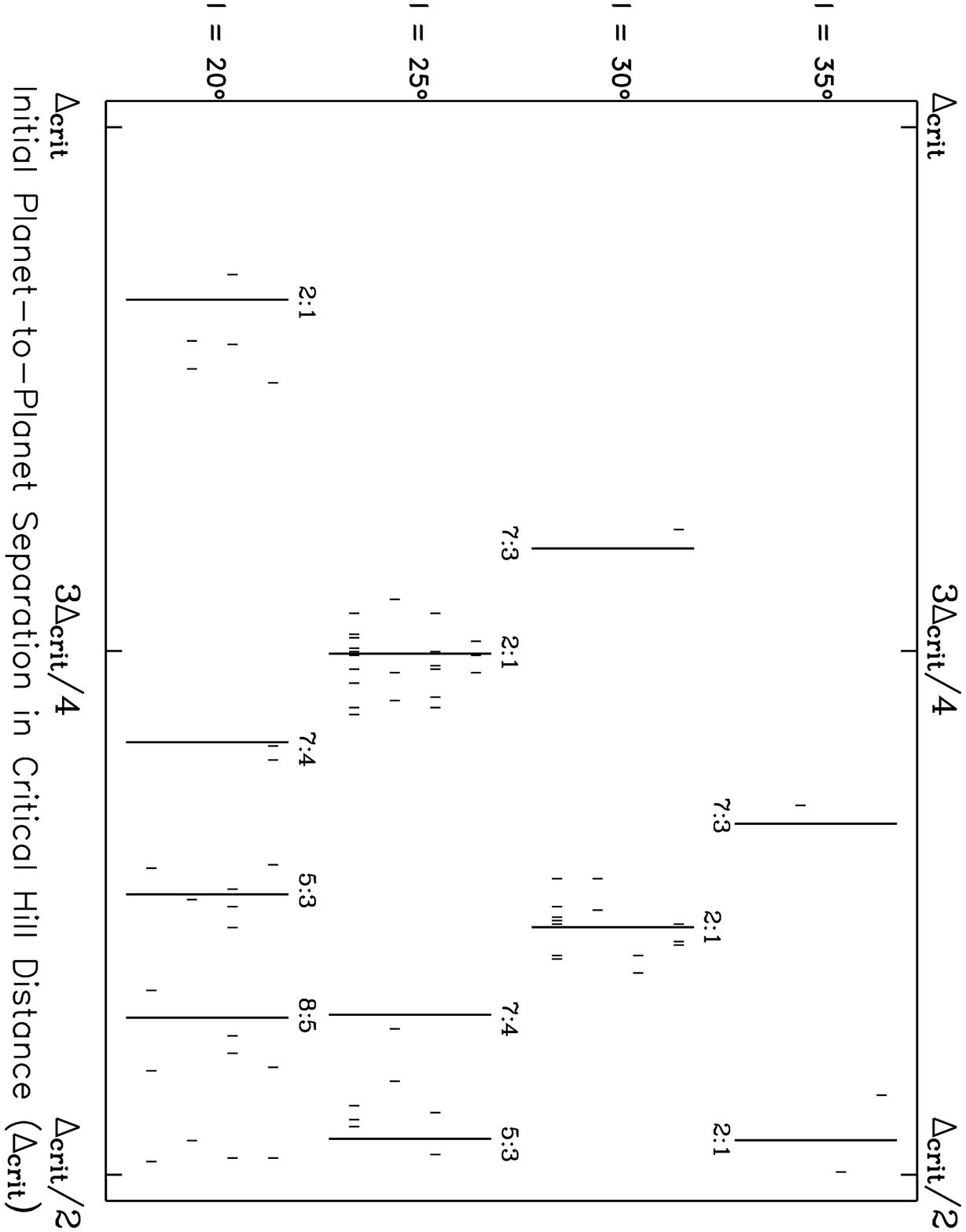,width=7.5truein,height=9.0truein}}
\caption{}
\label{fig7}
\end{figure}

Figure \ref{fig7}: Continuation of Fig. \ref{fig5}, with the locations of the
relevant major resonances shown, including the fourth-order 7:3 resonance.

\bigskip

\begin{figure}

\centerline{\psfig{figure=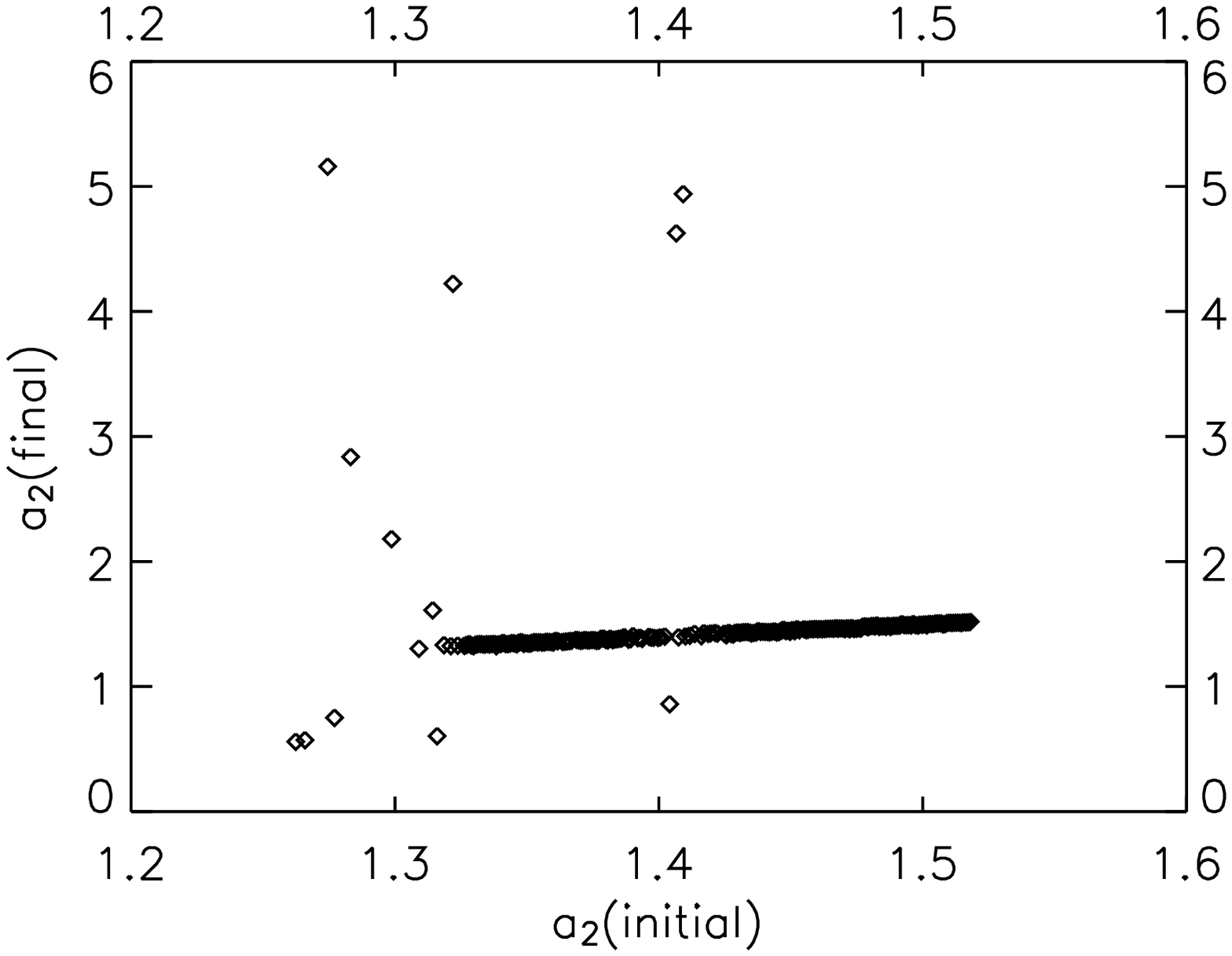,width=7.0truein,height=3.0truein}}
\centerline{}
\centerline{}
\centerline{\psfig{figure=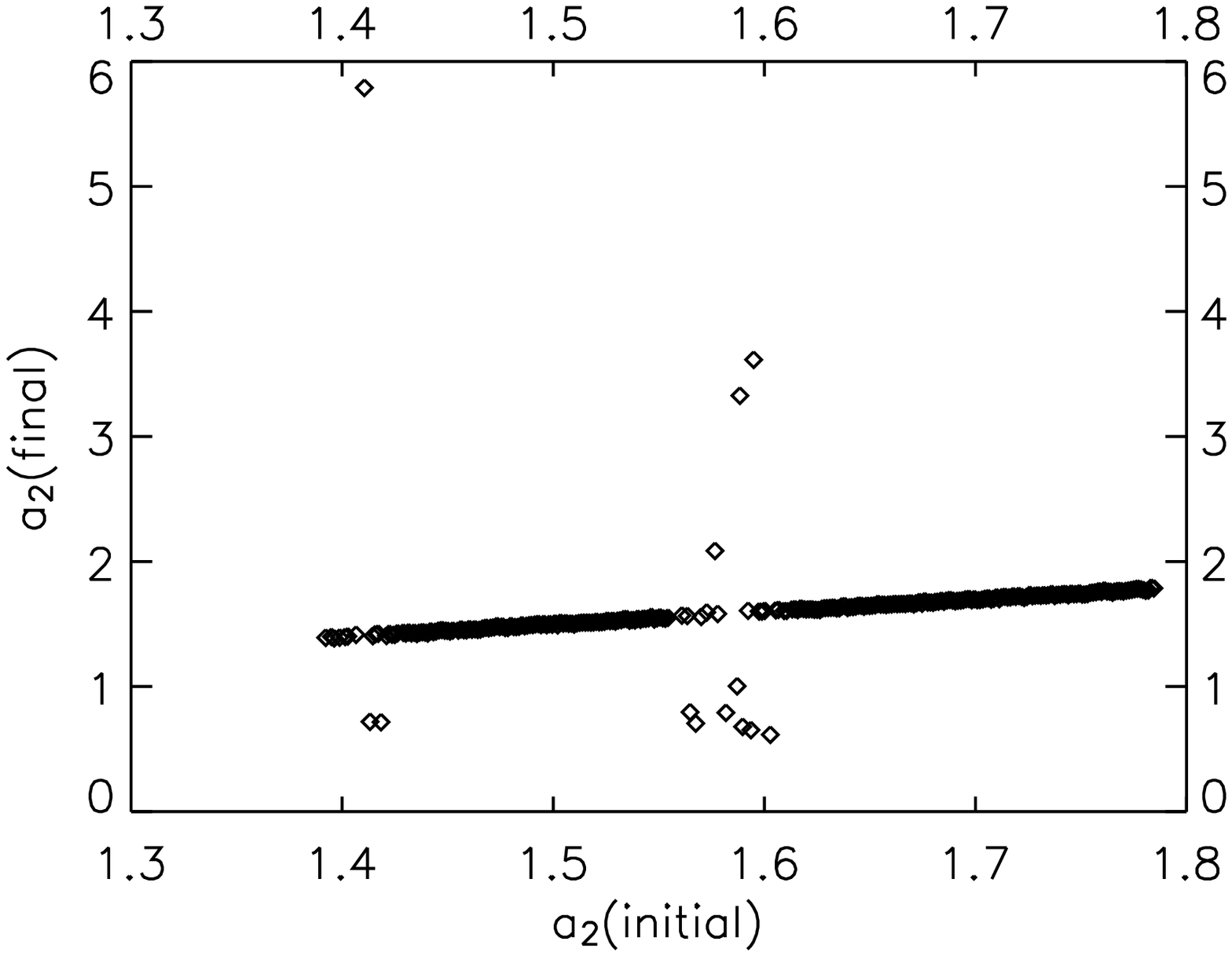,width=7.0truein,height=3.0truein}}
\centerline{}
\centerline{}
\centerline{\psfig{figure=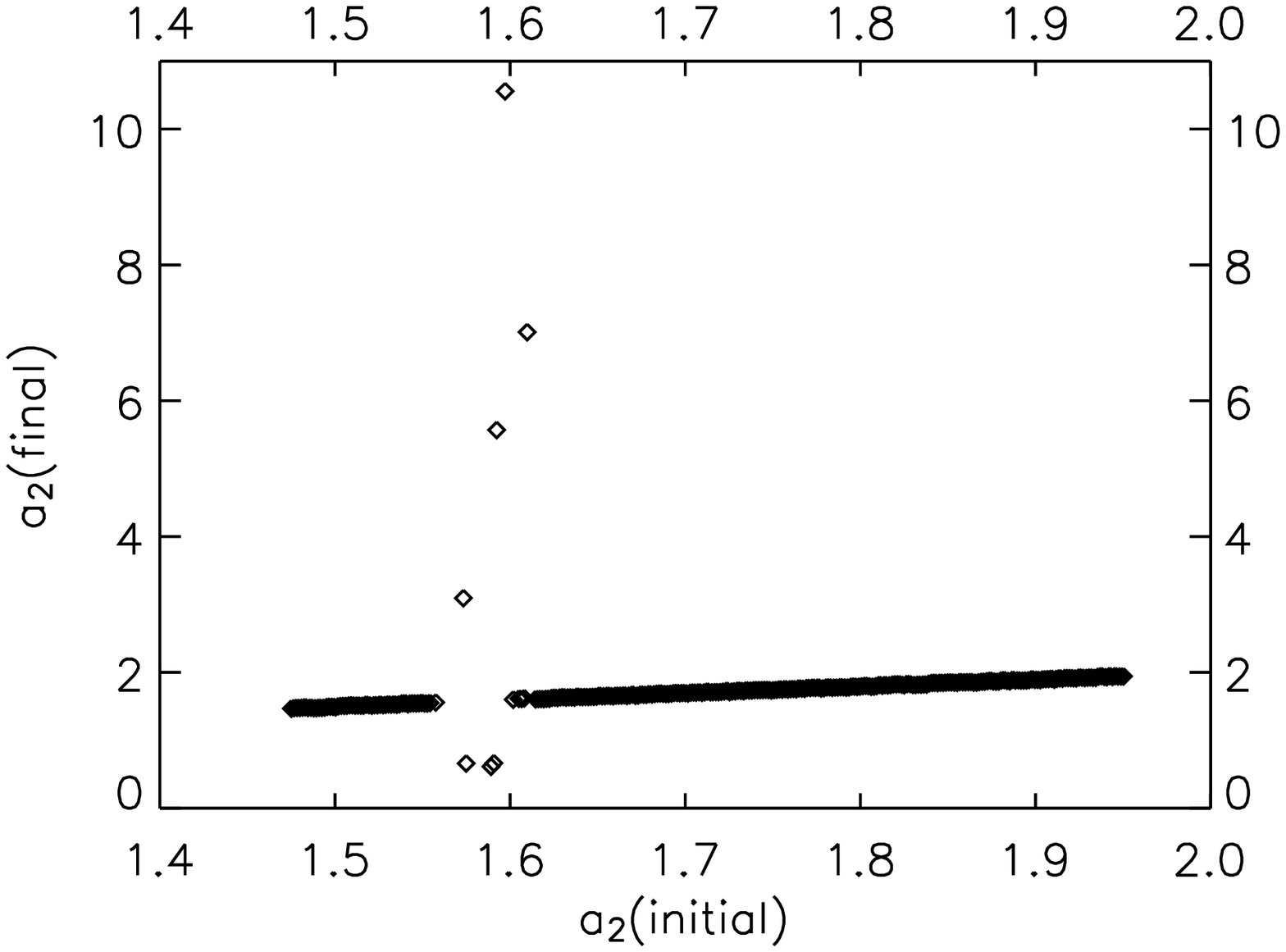,width=7.0truein,height=3.0truein}}
\caption{}
\label{fig8}
\end{figure}

Figure \ref{fig8}: Final vs. initial semimajor axis of the outer planet for the 
stable systems with an initial relative inclination of $15^{\circ}$ (upper panel),
$25^{\circ}$ (middle panel), and $30^{\circ}$ (lower panel).

\bigskip

\begin{figure}

\centerline{\psfig{figure=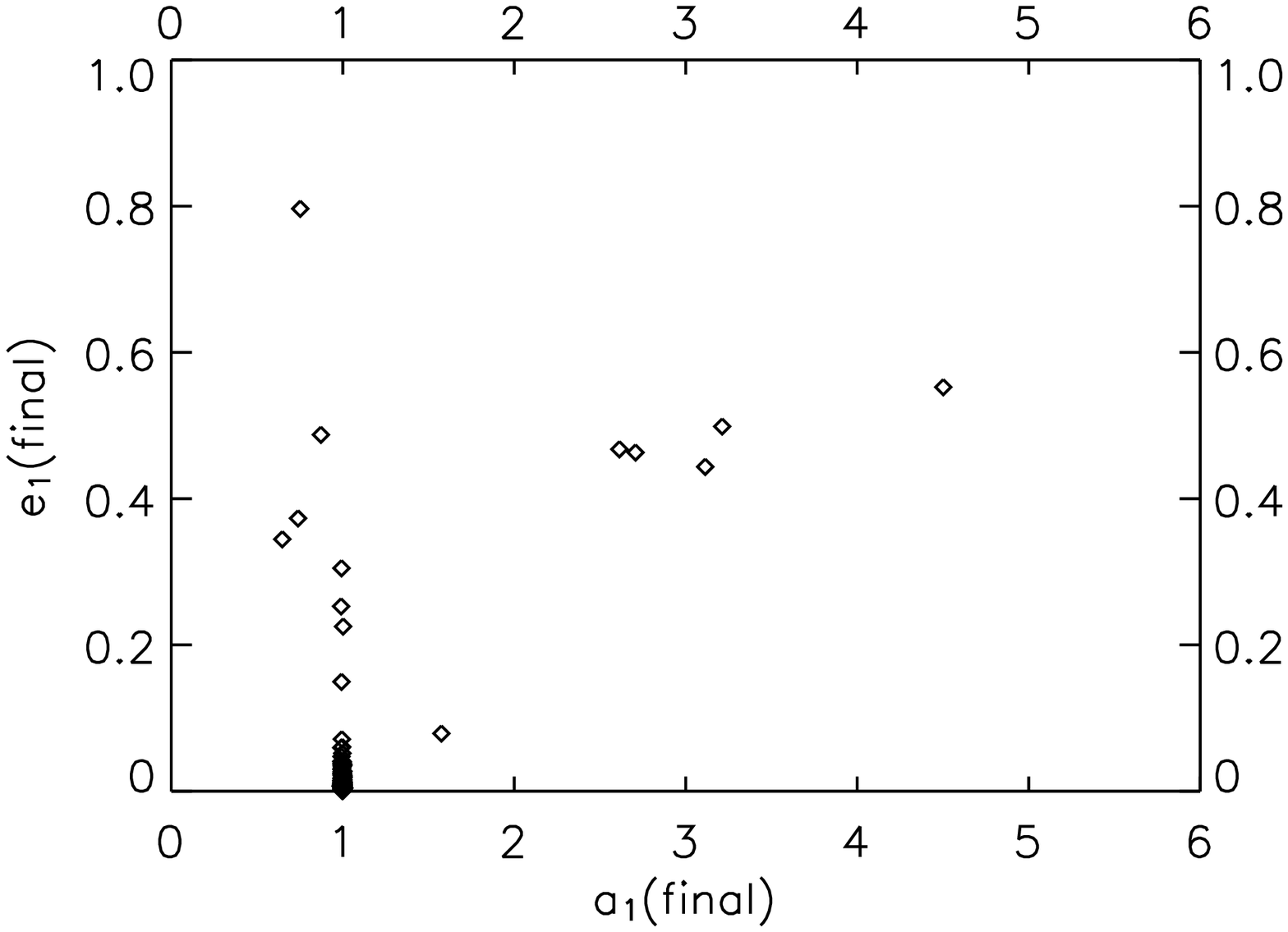,width=7.0truein,height=4.0truein}}
\centerline{}
\centerline{}
\centerline{\psfig{figure=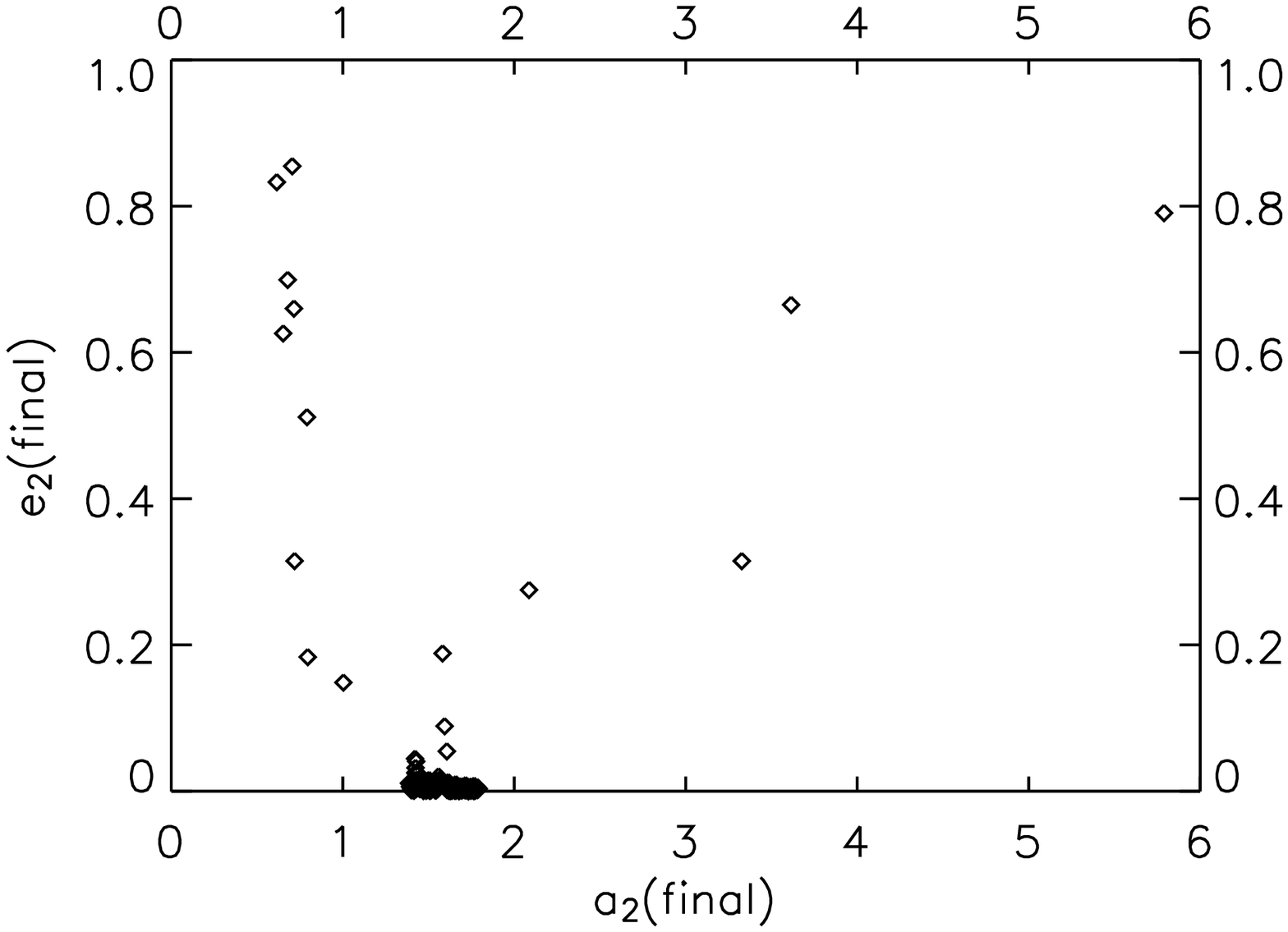,width=7.0truein,height=4.0truein}}
\caption{}
\label{fig9}
\end{figure}

Figure \ref{fig9}: Final values of eccentricity vs. semimajor axis for the inner
planet (top) and outer planet (bottom) in the systems from the middle panel
of Fig. \ref{fig8}.

\bigskip

\begin{figure}

\centerline{\psfig{figure=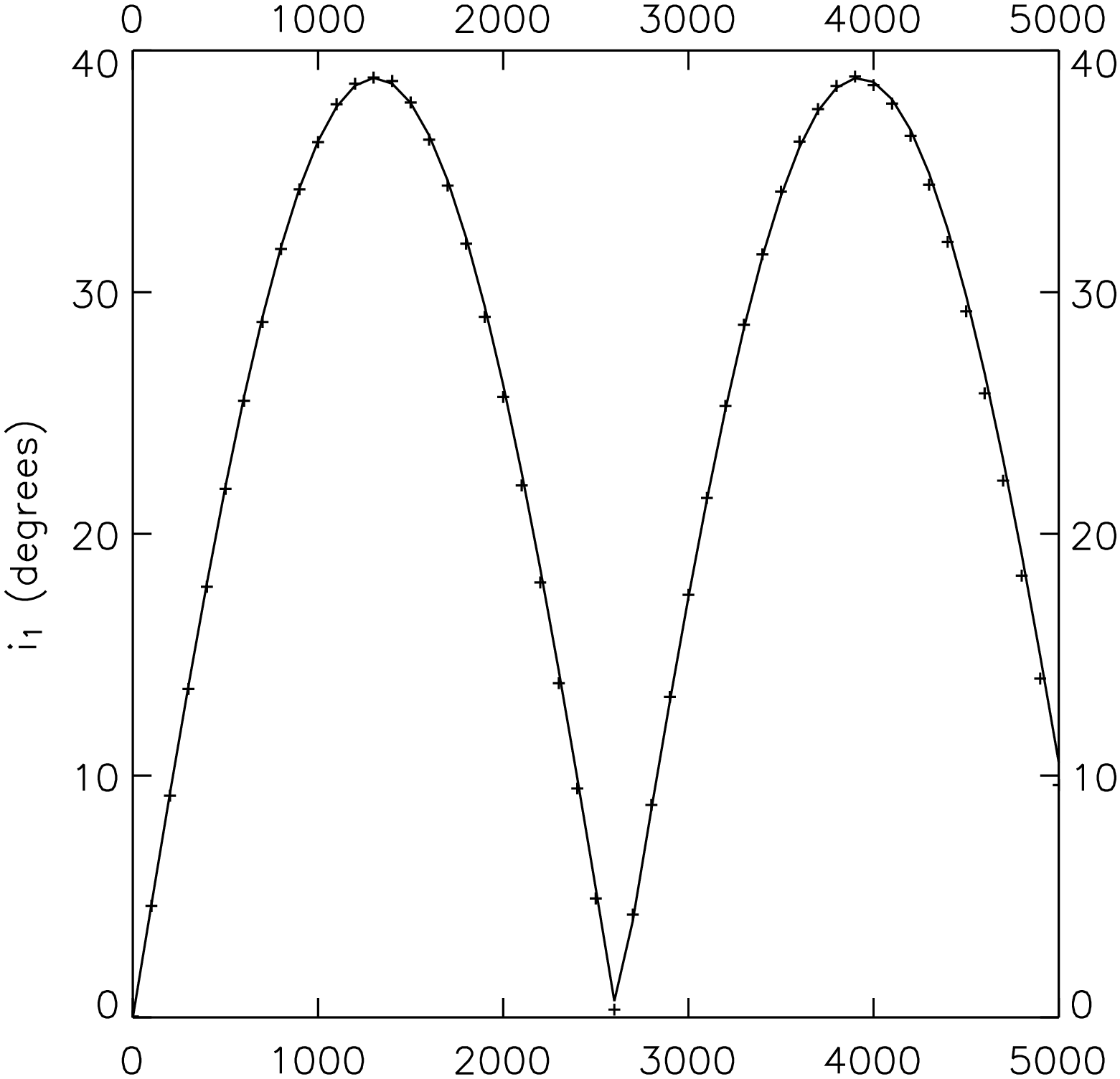,width=7.0truein,height=4.0truein}}
\centerline{}
\centerline{\psfig{figure=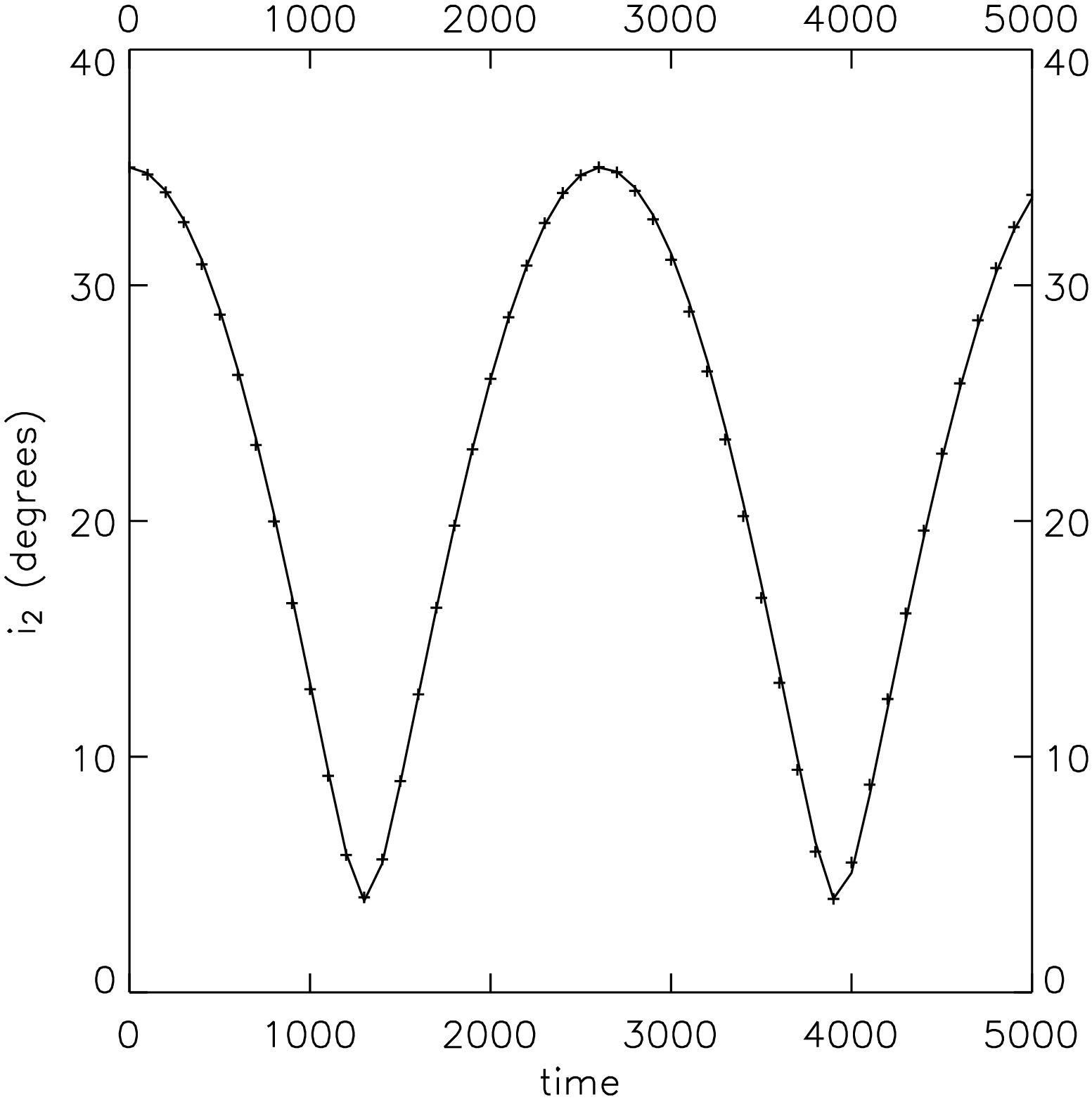,width=7.0truein,height=4.0truein}}
\caption{}
\label{fig10}
\end{figure}

Figure \ref{fig10}: Comparison of outputted data from numerical simulations (crosses)
and from Eqs. (\ref{i2t})-(\ref{i1t}) (solid lines) for a high initial
relative inclination ($35^{\circ}$) system with an initial outer semimajor axis
value only $0.0003$ away from instability.

\bigskip

\begin{figure}

\centerline{\psfig{figure=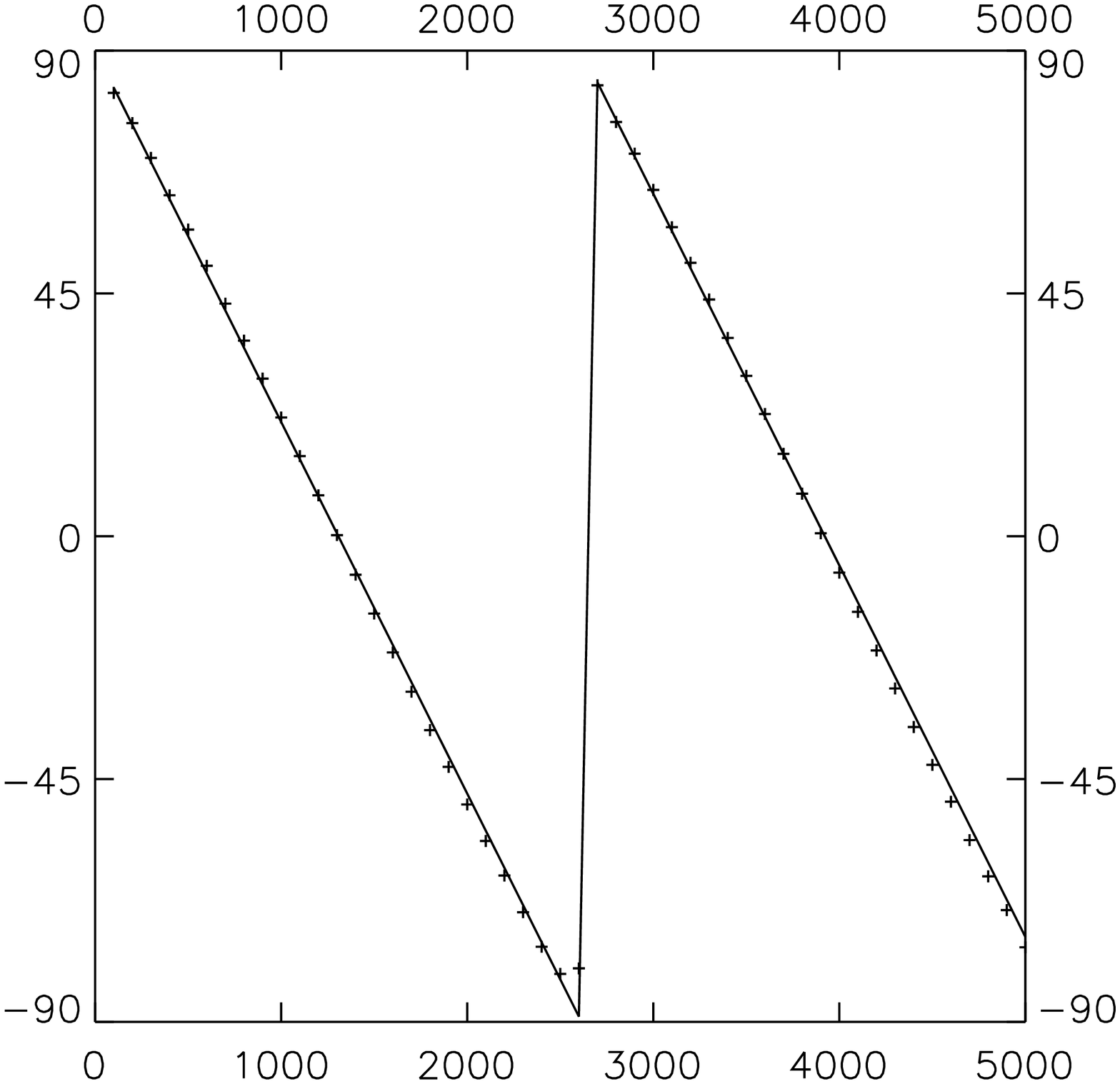,width=7.0truein,height=4.0truein}}
\centerline{}
\centerline{\psfig{figure=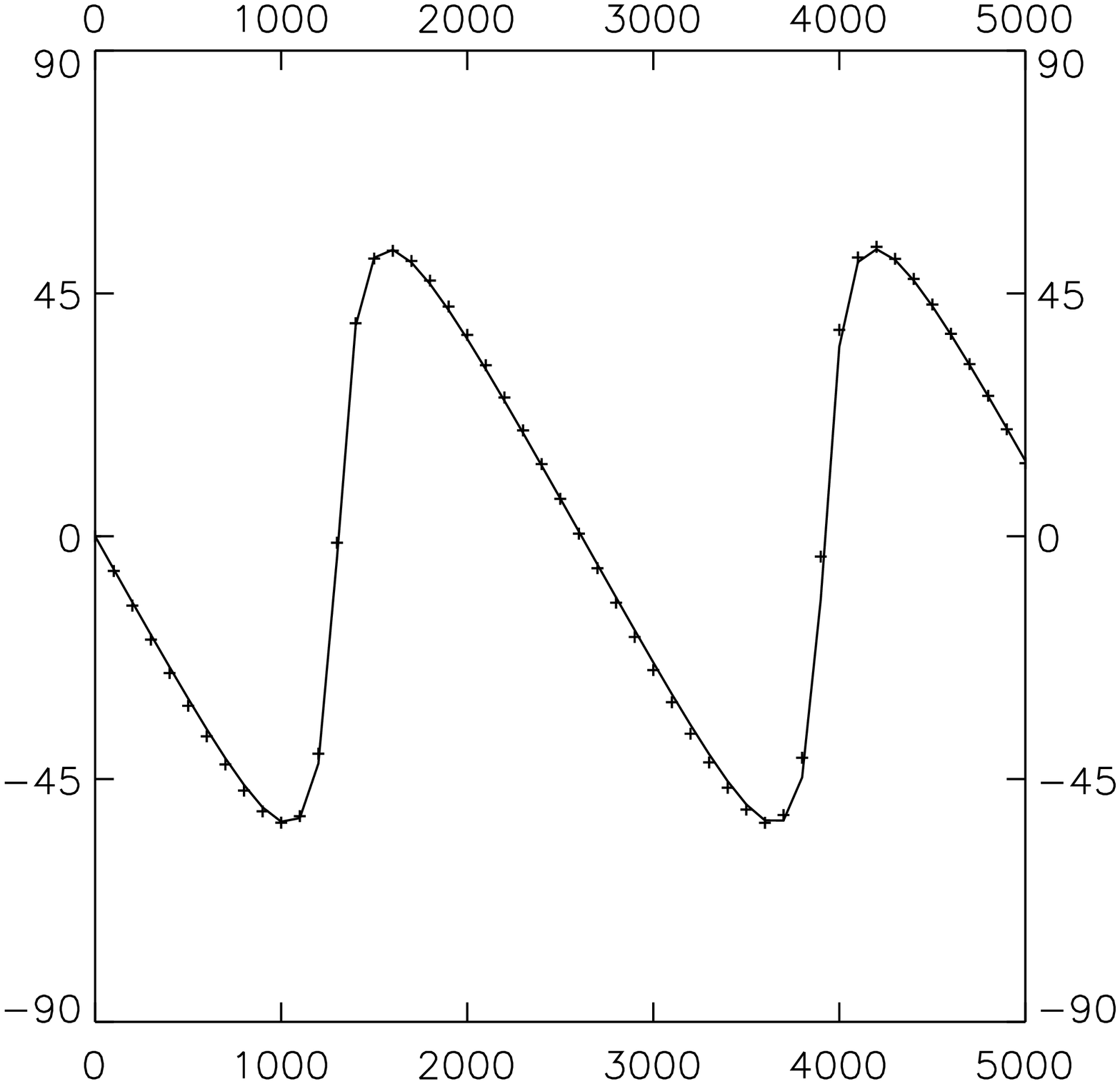,width=7.0truein,height=4.0truein}}
\caption{}
\label{fig11}
\end{figure}

Figure \ref{fig11}: Comparison of outputted data from numerical simulations (crosses)
and from Eqs. (\ref{node1var}) and (\ref{node2var}) (solid lines) for a high
inclination ($35^{\circ}$) system with an initial outer semimajor axis
value only $0.0003$ away from instability.

\bigskip

\begin{figure}

\centerline{\psfig{figure=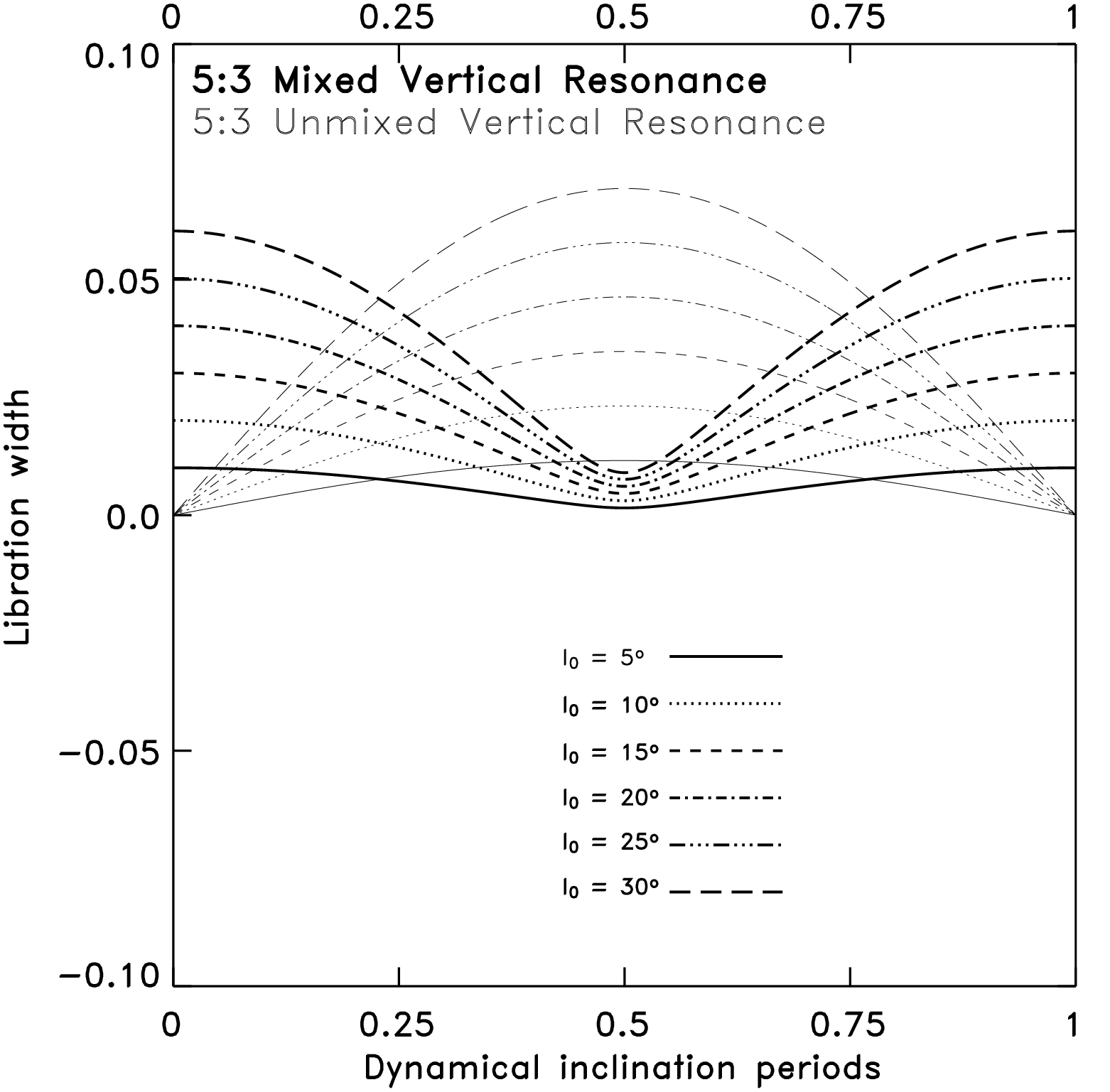,width=7.0truein,height=9.0truein}}
\caption{}
\label{fig12}
\end{figure}

Figure \ref{fig12}: Variation of libration width over one dynamical inclination period
for two different types of $5$:$3$ resonances and $6$ different relative
initial inclinations.

\bigskip

\begin{figure}

\centerline{\psfig{figure=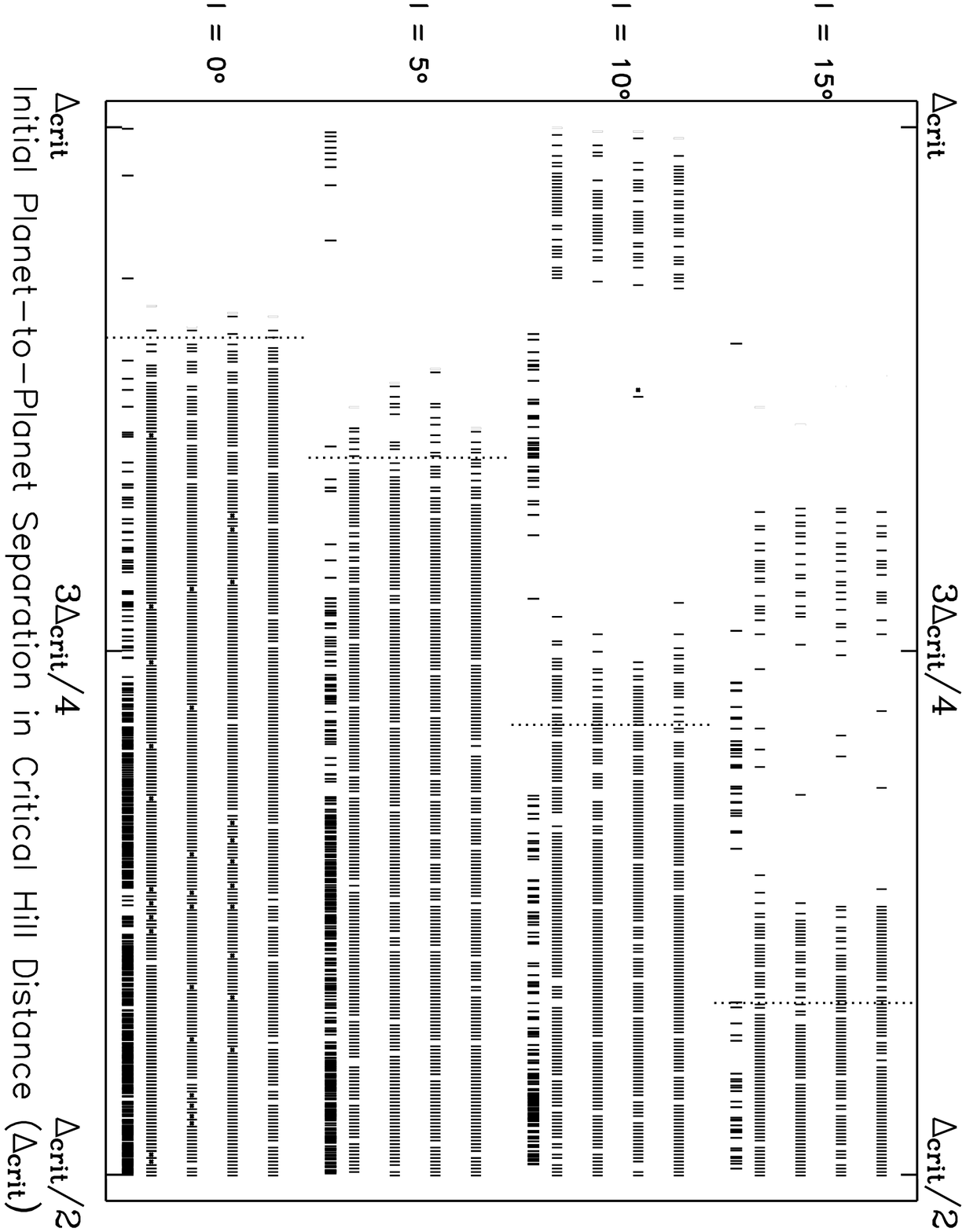,width=7.0truein,height=9.0truein}}
\caption{}
\label{fig13}
\end{figure}

Figure \ref{fig13}: Reproduction of Fig. \ref{fig4} with results of the second-order
nonplanar eccentric map superimposed.  The additional, darker bars represent
unstable systems from the map.

\bigskip

\begin{figure}

\centerline{\psfig{figure=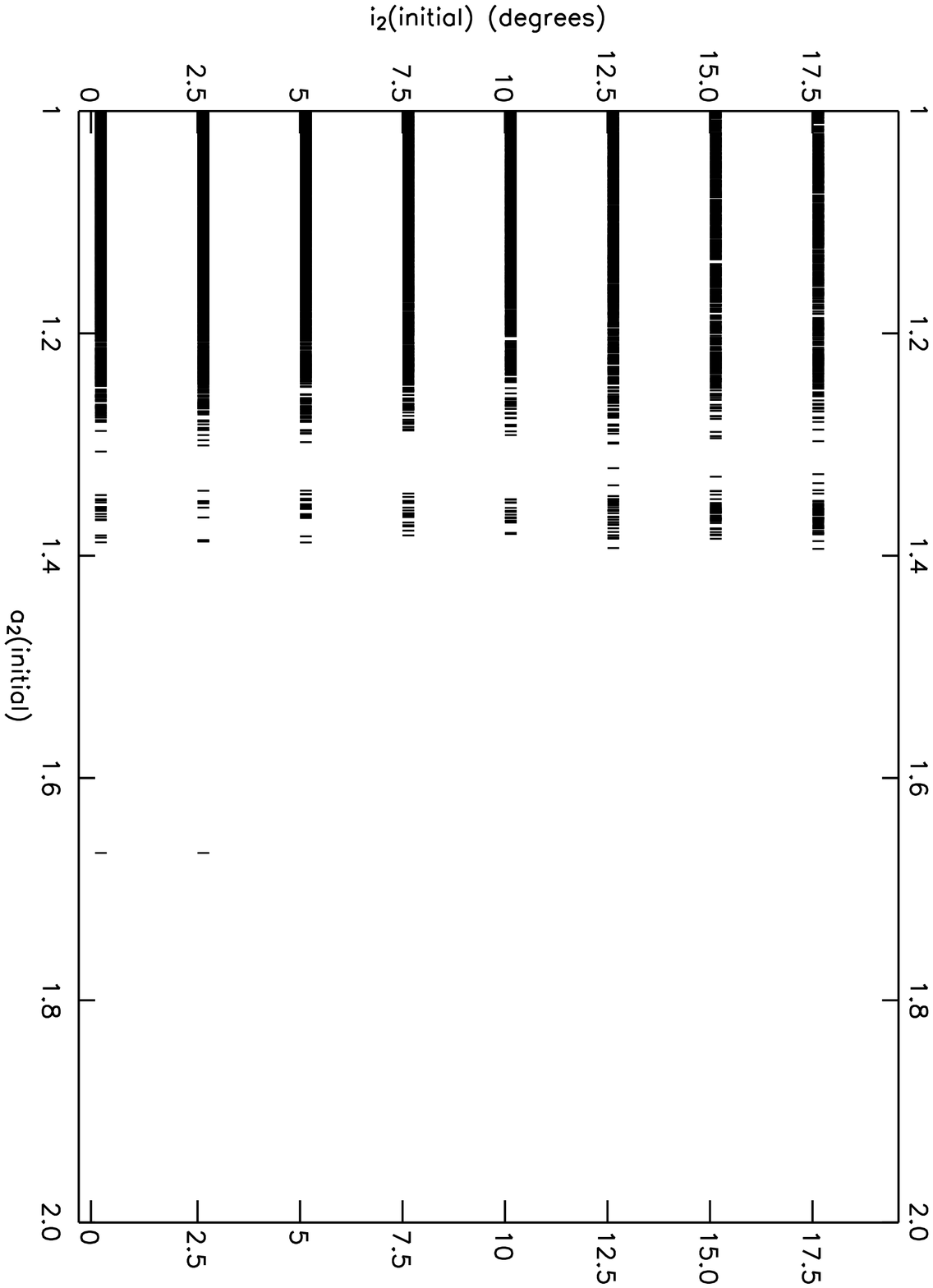,width=7.0truein,height=9.0truein}}
\caption{}
\label{fig14}
\end{figure}

Figure \ref{fig14}: Stability of planetary systems in the initial $a_2$ interval
[$1$, $2$], for inclinations up to $17.5^{\circ}$.  Each dark bar
represents an unstable system.

\bigskip

\begin{figure}

\centerline{\psfig{figure=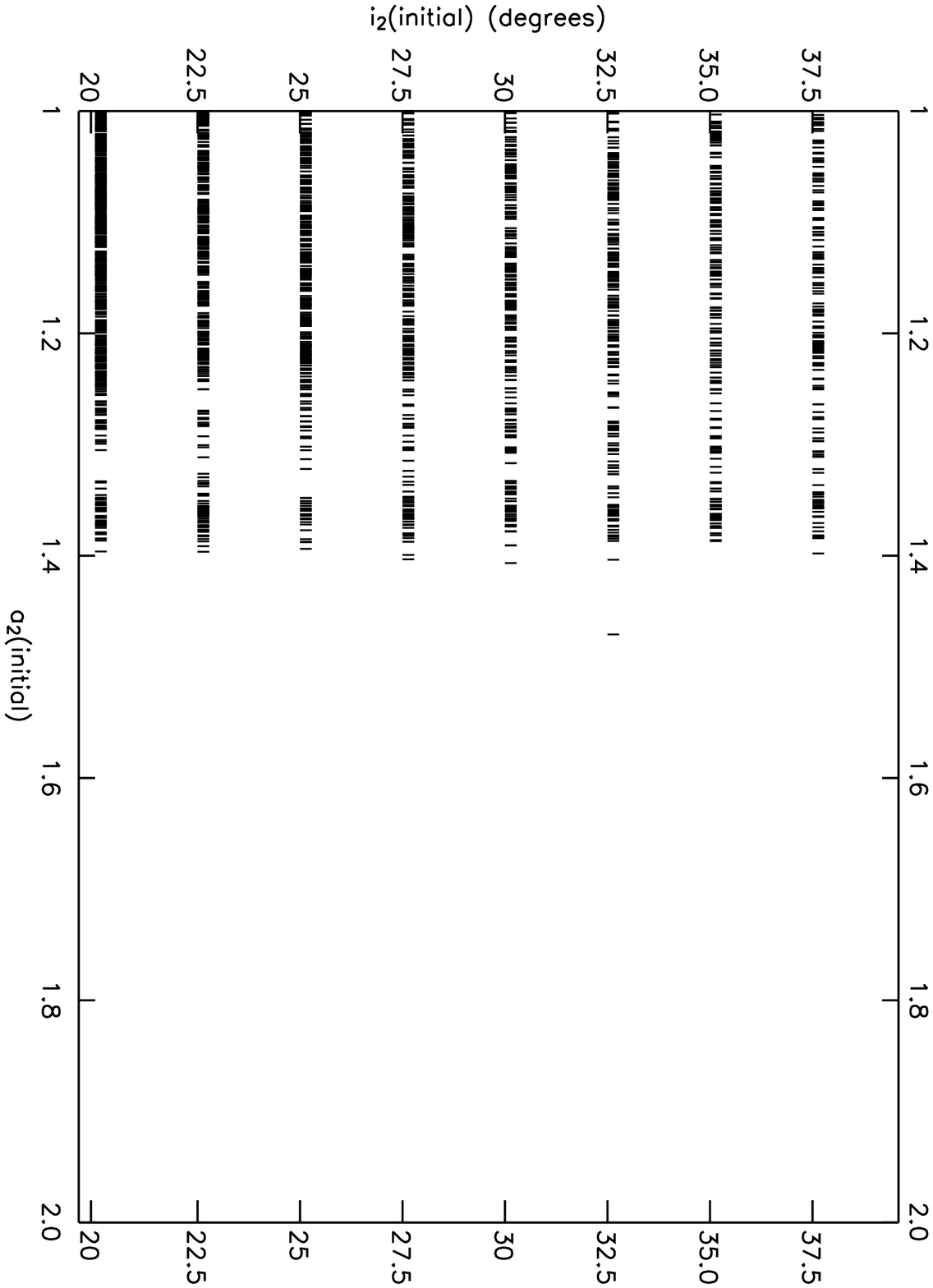,width=7.0truein,height=9.0truein}}
\caption{}
\label{fig15}
\end{figure}

Figure \ref{fig15}: Same as Fig. \ref{fig14}, except for inclinations in the range
[$20^{\circ}$, $37.5^{\circ}$].

\newpage

\bigskip
\bigskip

\newpage

\newpage

\newpage

\newpage

\newpage

\newpage

\newpage

\newpage

\newpage

\newpage

\newpage

\newpage

\newpage

\end{document}